\documentclass[a4paper,fleqn,usenatbib]{mnras}
\usepackage[T1]{fontenc}
\usepackage{ae,aecompl}
\usepackage{amsmath}
\usepackage{amsfonts}
\usepackage{amssymb}
\usepackage{graphicx}
\usepackage{color,xcolor}
\def\mnras{MNRAS}
\def\aap{A$\&$A}

\def\apj{ApJ}
\def\apjl{ApJ Letters}
\def\apjs{ApJS}
\def\pasp{PASP}
\def\aj{AJ}
\def\icarus{Icarus}

\newcommand{\msun}{\, {\rm M}_{\odot}}
\newcommand{\msunyr}{\, {\rm M}_{\odot}\,{\rm yr^{-1}}}
\newcommand{\rsun}{\, {\rm R}_{\odot}}
\newcommand{\au}{\, {\rm au}}

\title[External photoevaporation by OB runaways]{The observable impact of runaway OB stars on protoplanetary discs}
\author[G. A. L. Coleman et al]{Gavin A. L. Coleman$^{1}$\thanks{Email: gavin.coleman@qmul.ac.uk}, Jinyoung Serena Kim$^{2}$, Thomas J. Haworth$^{1}$, Peter A. Hartman$^{2,3}$ and
\newauthor Taylor C. Kalish$^{2}$\\
$^1$Astronomy Unit, Department of Physics and Astronomy, Queen Mary University of London, Mile End Road, London, E1 4NS, UK\\
$^2$Steward Observatory, University of Arizona, 933 N. Cherry Ave, Tucson, AZ 85721-0065, USA\\
$^3$Dept. of Astronomy, University of Massachusetts, Amherst, MA 01003, USA
}
\date{Accepted 2025 May 11. Received 2025 May 6; in original form 2025 February 10}
\pubyear{2025}
\begin{document}
\label{firstpage}
\pagerange{\pageref{firstpage}--\pageref{lastpage}}
\maketitle
\begin{abstract}
UV radiation from OB stars can drive ``external'' photoevaporative winds from discs in clusters, that have been shown to be important for disc evolution and planet formation. However, cluster dynamics can complicate the interpretation of this process. A significant fraction of OB stars are runaways, propagating at high velocity which might dominate over the wider cluster dynamics in setting the time variation of the UV field in part of the cluster. We explore the impact of a runaway OB star on discs and the observational impact that may have. We find that discs exposed to even short periods of strong irradiation are significantly truncated, and only rebound slightly following the ``flyby'' of the UV source. This is predicted to leave an observable imprint on a disc population, with those downstream of the OB star vector being more massive and extended than those upstream. Because external photoevaporation acts quickly, this imprint is less susceptible to being washed out by cluster dynamics for faster runaway OB stars. The Gaia proper motion vector of the B star 42 Ori in NGC 1977 is transverse to the low mass stellar population and so may make a good region to search for this signature in resolved disc observations. 
\end{abstract}

\section{Introduction}
\label{sec:intro}
The study of planet formation from circumstellar discs around young stars is increasingly being considered in the wider context of the star forming environment. External factors potentially impacting planet formation and evolution include enrichment by short-lived radionuclides \citep[e.g.][]{2016Icar..274..350L, 2024arXiv241017163E}, stellar flybys/runaways \citep[e.g.][]{1993MNRAS.261..190C, 2023EPJP..138...11C}, extended periods of accretion from the natal cloud \citep[e.g.][]{2020NatAs...4.1158P, 2024A&A...683A.133G, 2024arXiv240917220W} and, if there are nearby OB stars, their strong UV fields drive winds from discs in a process referred to as ``external photoevaporation'' \citep[a recent review is given by][]{Winter23} where the term distinguishes it from internal photoevaporation by the host star of the disc \citep[e.g.][]{2023ASPC..534..567P}. 

External photoevaporation now has a significant theoretical framework, describing the expected mass loss rates \citep{Johnstone98, Adams2004, Haworth18, Haworth23} and the associated effect on disc evolution \citep[e.g.][]{Clarke2007, 2018MNRAS.475.5460H, ConchaRamirez19, ColemanHaworth20,Coleman22, Coleman24MHD, Coleman25} and even planet formation \citep{Winter22, Qiao23, Huang24, Hallatt2025}. There is also empirical observational constraints on the mass loss rates, as well as probes of the physical conditions in the wind and ionisation front \citep[e.g.][]{1998ApJ...502L..71S, 1999ApJ...515..669S, Henney1999, MesaDelgado12, 2013MNRAS.430.3406T, Haworth21, Ballering23, Ballabio23, 2024Sci...383..988B, Aru24},  that both appear to be in good agreement with theoretical estimates. 

One of the key outstanding issues is that of the prevalence and time-scale of external photoevaporation \citep[see][for a detailed review of the other main issues]{PlanetFormationCollaboration}. It is difficult to gauge what fraction of discs are irradiated, by what strength of radiation field, and for how long. This is difficult because stellar clusters are complicated time-evolving systems, with stars moving in and out of high UV parts of the cluster over time \citep{2011PASP..123...14H, Qiao22, ParkerReiter22, Wilhelm23}. It is a problem that is also sensitive to the presence and dispersal of the natal molecular cloud \citep{Qiao22, Wilhelm23}. Overall this means that while possible statistical imprints of external photoevaporation on disc properties have been observed \citep[e.g.][]{Ansdell17, Eisner18, VanTerwsiga2020, VanTerwisga23, 2023A&A...680C...1M} they can be difficult to interpret \citep[e.g.][]{2021ApJ...913...95P}. Similarly, while individual discs undergoing external photoevaporation have been observed it is not clear how long that has been the case, especially given the short time-scale required to deplete a disc subjected to external photoevaporation \citep[e.g.][]{Henney1999, Winter19,Haworth21}. Disentangling the cluster dynamics remains a significant challenge. 

One interesting unexplored opportunity is that a significant number ($\sim20\%$) of OB stars end up as runaway stars early on in their lifetimes \citep[e.g.][]{2011Sci...334.1380F}. These are ejected from their cluster due to binary interactions and can travel with velocities up to many tens, or even hundreds, of km\,s$^{-1}$ \citep[for a recent catalogue see][]{2025arXiv250202658C}. Note that slower moving stars ejected from a cluster are referred to as walkaways and are potentially more numerous than the faster runaways \citep[e.g.][]{2019A&A...624A..66R}. These significant UV sources propagating out of their birth cluster, or through another one, has the potential to leave a well characterised time varying UV imprint on a cluster and constrains the time-scale that any given star/disc is exposed to very strong UV irradiation. Such systems could provide a new insight into the way external photoevaporation itself operates, and could simultaneously modify the expected fraction of discs in clusters that are subject to external photoevaporation. 

In this paper we introduce models of star/disc systems subjected to the time-varying UV field of a passing runaway or walkaway OB star. We study the impact of this short burst of UV irradiation on the disc, and how the properties of flyby events influence the evolution of protoplanetary discs. We also study the possible observational imprints of this scenario on the disc radii in clusters, suggesting that such an imprint may be observable in the cluster NGC 1977 due to the B star 42 Ori moving transverse to the low mass stellar population.

\section{Physical Model}
\label{sec:base_model}

Protoplanetary discs lose mass through accretion on to the central star and by photoevaporative winds launched from the surface layers of the disc. Typically, the outer edges of protoplanetary discs are regulated through the balance of viscous spreading and the loss of disc material through external photoevaporation \citep{Coleman22,Coleman24MHD}. However, most studies assume that such discs are evolving in isolation, or in constant FUV radiation fields, and so do not take into account the movement of stars throughout a cluster. Though, recent work has explored the effects of shielding on the evolution of protoplanetary discs \citep{Qiao22,Qiao23}, whilst other work has explored the effects of stellar clusters on the evolution of protoplanetary disc fractions \citep{ConchaRamirez19,ConchaRamirez21}. Since stars mutually gravitationally interact, they will attain relative velocities to each other that will allow them to move closer to the other stars and their inherent discs. Such flybys, if they involve massive stars, will result in much stronger external radiation fields that can significantly affect the evolution of protoplanetary discs. Below, we will describe how we model the evolution of protoplanetary discs, including prescriptions for internal and external photoevaporation, as well as the effects of a flyby from a massive star, which will be the main focus of this paper. 

\subsection{Gas disc}
For the underlying gas disc model, we use the model outlined in \citet{Coleman21} that uses a 1D viscous disc model\footnote{An MHD-wind driven model should yield similar evolutionary outcomes with appropriate values for such models, since photoevaporation is found to dominate the mass-loss mechanisms from discs which determines their pathway of evolution\citep{Coleman24MHD}.} where the evolution of the gas surface density $\Sigma$ is solved through the standard diffusion equation.

\begin{equation}
    \dot{\Sigma}(r)=\dfrac{1}{r}\dfrac{d}{dr}\left[3r^{1/2}\dfrac{d}{dr}\left(\nu\Sigma r^{1/2}\right)\right]-\dot{\Sigma}_{\rm PE}(r)
\end{equation}
where $\nu=\alpha H^2\Omega$ is the disc viscosity with viscous parameter $\alpha$ \citep{Shak}, $H$ the disc scale height, $\Omega$ the Keplerian frequency, and $\dot{\Sigma}_{\rm PE}(r)$ is the rate of change in surface density due to photoevaporative winds.
Following \citet{Coleman22} we include EUV and X-ray internal photoevaporative winds from the central star (detailed in section \ref{sec:internalPhoto}) as well as winds launched from the outer disc by far ultraviolet (FUV) radiation emanating from nearby massive stars (e.g. O-type stars, see section \ref{sec:externalPhoto}).
We assume that the photoevaporative mass loss rate at any radius in the disc is the maximum of the internally and externally driven rates 
\begin{equation}
    \dot{\Sigma}_{\rm PE}(r) ={\rm max}\left(\dot{\Sigma}_{\rm I,X}(r),\dot{\Sigma}_{\rm E,FUV}(r)\right)
\end{equation}
where the subscripts I and E refer to contributions from internal and external photoevaporation.

The disc is assumed to be in thermal equilibrium, and so the temperature is calculated by balancing irradiation heating from the central star, background heating from the residual molecular cloud, viscous heating and blackbody cooling. To attain thermal equilibrium, we follow \citet{Coleman21} and use an iterative method to solve the following equation \citep{Dangelo12}
\begin{equation}
    Q_{\rm irr} + Q_{\nu} + Q_{\rm cloud} - Q_{\rm cool} = 0
\end{equation}
where $Q_{\rm irr}$ is the radiative heating rates due to the central star, $Q_{\nu}$ is the viscous heating rate per unit area of the disc, $Q_{\rm cloud}$ is the radiative heating due to the residual molecular cloud, and $Q_{\rm cool}$ is the radiative cooling rate.

\subsection{Internal Photoevaporation}
\label{sec:internalPhoto}
The absorption of high energy radiation from the host star by the disc can heat the gas above the local escape velocity, and drive internal photoevaporative winds. EUV irradiation creates a layer of ionised hydrogen with temperature $\sim$10$^4$~K \citep{Clarke2001}, whilst X-rays penetrate deeper into the disc and are capable of heating the gas up to around $\sim$10$^4$~K \citep{Owen10} and so for low mass stars X-rays are expected to generally dominate over the EUV for setting the mass loss rate. FUV radiation penetrates deeper still, creating a neutral layer of dissociated hydrogen with temperature of roughly 1000K \citep{Matsuyama03}. The overall interplay between the EUV, FUV and X-rays is a matter of ongoing debate.  \cite{Owen12} find that including the FUV heating simply causes the flow beneath the sonic surface to adjust, but otherwise retains similar mass loss rates. However other works suggest a more dominant role of the FUV \citep[e.g.][]{Gorti09,Gorti15}. Recent models including all three fields suggest a more complicated interplay \citep[e.g.][]{Wang17, Nakatani18,Ercolano21}. Additionally, the outcome also depends sensitively on how the irradiated spectrum is treated \citep{Sellek22}. 

The radiation hydrodynamic models of \cite{Owen12} used pre-computed X-ray driven temperatures as a function of the ionisation parameter ($\xi = L_X / n /r^2$) wherever the column to the central star is less than $10^{22}$cm$^{-2}$ (thus optically thin). More recently this approach has been updated with a series of column-dependent temperature prescriptions \citep{Picogna19,Picogna21,Ercolano21}.  

We follow \citet{Picogna21} who further build on the work of \cite{Picogna19} and \cite{Ercolano21} and find that the mass loss profile from internal X-ray irradiation is approximated by
\begin{equation}
\label{eq:sig_dot_xray}
\begin{split}
\dot{\Sigma}_{\rm I,X}(r)=&\ln{(10)}\left(\dfrac{6a\ln(r)^5}{r\ln(10)^6}+\dfrac{5b\ln(r)^4}{r\ln(10)^5}+\dfrac{4c\ln(r)^3}{r\ln(10)^4}\right.\\
&\left.+\dfrac{3d\ln(r)^2}{r\ln(10)^3}+\dfrac{2e\ln(r)}{r\ln(10)^2}+\dfrac{f}{r\ln(10)}\right)\\
&\times\dfrac{\dot{M}_{\rm X}(r)}{2\pi r} \dfrac{\msun}{\au^2 {\rm yr}}
\end{split}
\end{equation}
where
\begin{equation}
\label{eq:m_dot_r_xray}
    \dfrac{\dot{M}_{\rm X}(r)}{\dot{M}_{\rm X}(L_{X})} = 10^{a\log r^6+b\log r^5+c\log r^4+d\log r^3+e\log r^2+f\log r+g}.
\end{equation}
Recently \citet{Sellek24} provided updated estimates for $a$--$g$, by including the effects of additional cooling as a result of the excitation of O by neutral H. They found that $a=-0.6344$, $b=6.3587$, $c=-26.1445$, $d=56.4477$, $e=-67.7403$, $f=43.9212$, and $g=-13.2316$.

We follow \citet{Komaki23} and apply a simple approximation to the outer regions of the disc where the internal photoevaporation rates drop to zero.
As the temperature of X-ray irradiated gas varies from $\sim 10^2$--$10^4$ K depending on the distance in the disc \citep[e.g.][]{Owen10}, we conservatively define the radius at which the internal photoevaporation scheme drops off as the gravitational radius for $10^3$ K gas.
We therefore apply the following approximation at radial distances greater than $r_{\rm rgx}$\footnote{Note that we use $r_{\rm rgx}$ as an arbitrary value, and a different radius in the disc would also be suitable as long as it is sufficiently distant from where internal photoevaporation mostly operates.}
\begin{equation}
    \dot{\Sigma}_{\rm I,X,ap} = \dot{\Sigma}_{\rm rgx}\left(\frac{r}{r_{\rm rgx}}\right)^{-1.578}
\end{equation}
where $\dot{\Sigma}_{\rm rgx}$ is equal to eq. \ref{eq:sig_dot_xray} calculated at $r=r_{\rm rgx}$, and
\begin{equation}
    r_{\rm rgx} = \dfrac{GM_*}{c_{\rm s}^2}
\end{equation}
where $c_{\rm s}$ is the sound speed for gas of temperature $T=10^3 K$, and $\mu=0.5$. The exponent of -1.578 is consistent with the gradients found in the mass-loss profiles observed in the outer regions of the disc before the cut-off is applied in \citet{Picogna21}.

In the outer regions of the disc the loss in gas surface density due to internal photoevaporation then becomes
\begin{equation}
    \dot{\Sigma}_{\rm I}(r) = \max(\dot{\Sigma}_{\rm I,X}(r),\dot{\Sigma}_{\rm I,X,ap})
\end{equation}

Following \cite{Ercolano21} the integrated mass-loss rate, dependent on the stellar X-ray luminosity, is given as
\begin{equation}
    \log_{10}\left[\dot{M}_{X}(L_X)\right] = A_{\rm L}\exp\left[\dfrac{(\ln(\log_{10}(L_X))-B_{\rm L})^2}{C_{\rm L}}\right]+D_{\rm L},
\end{equation}
in $\msunyr$, with $A_{\rm L} = -1.947\times10^{17}$, $B_{\rm L} = -1.572\times10^{-4}$, $C_{\rm L} = -0.2866$, and $D_{\rm L} = -6.694$. By including additional cooling effects due to the excitation of O from neutral H, \citet{Sellek24} found that the mass loss rates found in \citet{Picogna21} are overestimated by factor $\sim9$, and as such we apply this correction to the equation above.

\subsection{External Photoevaporation}
\label{sec:externalPhoto}

In addition to internal winds driven by irradiation from the host star, winds can also be driven from the outer regions of discs by irradiation from external sources. Massive stars dominate the production of UV photons in stellar clusters and hence dominate the external photoevaporation of discs. External photoevaporation has been shown to play an important role in setting the evolutionary pathway of protoplanetary discs \citep{Coleman22}, their masses \citep{Mann14,Ansdell17,VanTerwisga23}, radii \citep{Eisner18} and lifetimes \citep{Guarcello16,ConchaRamirez19,Sellek20} even in weak UV environments \citep{Haworth17}.
We do not include shielding of the protoplanetary discs, i.e. by the nascent molecular cloud, that has been shown to have an effect on the effectiveness of external photoevaporation \citep{Qiao22,Qiao23,Wilhelm23}, but will instead assume that the discs are evolving in environments with a weak base UV field strength.

In our simulations, the mass loss rate due to external photoevaporation is calculated by interpolating over the recently updated \textsc{fried} grid \citep{Haworth23}. The \textsc{fried} grid provides mass loss rates for discs irradiated by FUV radiation as a function of the star/disc/FUV parameters. In our simulations, we determine the mass loss rate at each time step by linearly interpolating \textsc{fried} in three dimensions: disc size $R_{d}$, disc outer edge surface density $\Sigma_{\textrm{out}}$ and FUV field strength $F_{\rm{UV}}$.

We evaluate the \textsc{fried} mass loss rate at each radius from the outer edge of the disc down to the radius that contains 80$\%$ of the disc mass. We choose this value as 2D hydrodynamical models show that the vast majority of the mass loss from external photoevaporation, comes from the outer 20\% of the disc \citep{Haworth19}. The change in gas surface density is then calculated as
\begin{equation}
    \dot{\Sigma}_{\textrm{ext, FUV}}(r) = G_{\rm sm} \frac{\dot{M}_\textrm{{ext}}(R_{\textrm{\textrm{max}}})}{\pi(R^2_\textrm{{d}} - {R_{\textrm{\textrm{max}}}}^2)+A_{\rm sm}}, 
\end{equation}
where $A_{\rm sm}$ is a smoothing area equal to 
\begin{equation}
A_{\rm sm} = \dfrac{\pi(R_{\rm max}^{22}-(0.1 R_{\rm max})^{22})}{11R_{\rm max}^{20}}
\end{equation}
and $G_{\rm sm}$ is a smoothing function
\begin{equation}
    G_{\rm sm} = \dfrac{r^{20}}{R_{\rm max}^{20}}.
\end{equation}

The {\sc fried} grid contains multiple subgrids that vary the PAH-to-dust ratio ($\rm f_{PAH}$) and specify whether or not grain growth has occurred.
The effects of using different combinations of these parameters will be explored in future work, but we do not expect such changes to affect the differences between viscous and MHD wind driven discs.
The combination we use here is $\rm f_{PAH}=1$ (an interstellar medium, ISM,-like PAH-to-dust ratio) and assume that grain growth has occurred in the outer disc, depleting it and the wind of small grains which reduces the extinction in the wind and increases the mass loss rate compared to when dust is still ISM-like. This combination of parameters results in PAH-to-gas abundances comparable to our limited observational constraints on that value \citep{Vicente13}. 

\subsection{Flyby of massive stars}
\label{sec:flyby_model}

The model for external photoevaporation above assumes that the FUV field strength remains constant over time. However, as massive O-type or B-type stars are the main providers for FUV radiation, and since we assume that these stars are moving through a cluster of stars, then their proximity to evolving protoplanetary discs would vary over time. To model this we introduce a temporally variable FUV field, whereby the FUV field strength smoothly rises from its background value to the maximum value when the massive star is at its closest approach. After the star has passed the disc, the FUV field strength smoothly reverts back down to the background value. This results in the FUV field strength being equal to
\begin{equation}
\label{eq:uv_implementation}
    F_{\rm UV} = F_{\rm UV, b} + (F_{\rm UV, max}-F_{\rm UV, b})\times \tanh{\left(\left(\dfrac{\tau_{\rm fb}}{t_{\rm fb}-t}\right)^{2}\right)}
\end{equation}
where $F_{\rm UV, b}$ is the base FUV field strength, and $F_{\rm UV, max}$ is the maximum FUV field strength depending on the properties of the star performing the flyby. The flyby time is denoted by $t_{\rm fb}$, taken at the time of closest approach / peak FUV intensity, whilst $\tau_{\rm fb}$ is the flyby time-scale.

\begin{figure}
\centering
\includegraphics[scale=0.5]{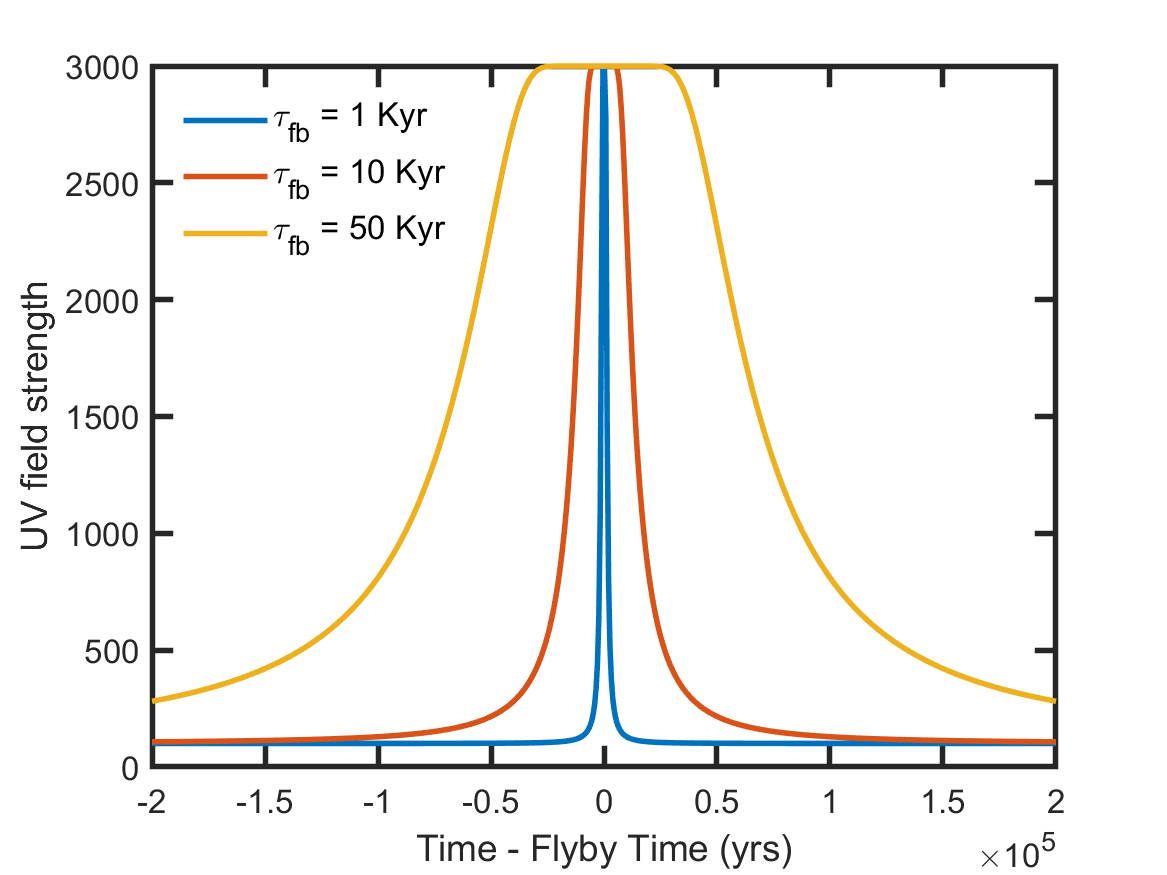}
\caption{Temporal evolution of the UV field centered around the flyby time for different flyby time-scales of: 1 Kyr (blue), 10 Kyr (red) and 50 Kyr (yellow). The profiles follow the equation outlined in Eq. \ref{eq:uv_implementation}.}
\label{fig:flyby_profile}
\end{figure}

In Fig. \ref{fig:flyby_profile} we show the change in the instantaneous UV field as a function of time before and after the flyby time. For this example, the FUV field strength increased from a base value of 100 G$_0$, up to a value of 3000 G$_0$. The different coloured lines denote different flyby time-scales, ranging from 1 Kyr (blue), to 10 Kyr (red) and 50 Kyr (yellow). The smooth increases to and from the maximum UV value are clearly evident here, even for the faster flyby time-scale.

\subsection{Simulation Parameters}
\label{sec:pop_parameters}

In this work we mainly explore the effects of flybys of massive stars on the evolution of protoplanetary discs. Therefore we do not vary the underlying disc parameters for our models. To account for the flyby star, we vary the time of the flyby in the disc's lifetime $t_{\rm fb}$, the time-scale of the flyby $\tau_{\rm fb}$, i.e. for how long the UV radiation field is increased over the ambient background field, and on the strength of the change in UV radiation $F_{\rm UV, max}$ to account for different mass flyby stars or greater proximity flyby events.

We initialise our disc following \citet{Lynden-BellPringle1974}
\begin{equation}
    \Sigma = \Sigma_0\left(\frac{r}{1\au}\right)^{-1}\exp{\left(-\frac{r}{r_{\rm c}}\right)}
\end{equation}
where $\Sigma_0$ is the normalisation constant set by the total disc mass, (for a given $r_{\rm c}$), and $r_{\rm c}$ is the scale radius, which sets the initial disc size, taken here to be equal to 50 $\au$. We take the initial disc mass to be equal to 0.1$\msun$.
Table \ref{tab:parameters} shows the simulation parameters that we used in this work.

\begin{table}
    \centering
    \begin{tabular}{c|c}
    \hline
   Constant Parameter & Value \\
    \hline
        $r_{\rm in} (\au)$ & 0.05 \\
        $r_{\rm out} (\au)$ & 500 \\
        $r_{\rm c} (\au)$ & 50 \\
        $M_{*} (\msun)$ & 1 \\
        $T_* (K)$ & 4300 \\
        $R_* (\rsun)$ & 2 \\
        $M_{\rm d} (\msun)$ & 0.1 \\
        $L_X (\log_{10}(\rm erg~s^{-1}))$ & 29.5 \\
        $\alpha$ & $10^{-3}$\\
        \hline
        Variable Parameter & Value \\
        \hline
        UV field ($\rm G_0$) & $10^1$--$10^5$ \\
        $t_{\rm fb}$ (Myr) & 0.1 -- 2\\
        $\tau_{\rm fb}$ (Kyr) & 1, 5, 10, 50, 100\\
    \hline
    \end{tabular}
    \caption{Simulation Parameters for the models discussed in this paper.}
    \label{tab:parameters}
\end{table}

\section{Results}
\label{sec:results}

With Sect. \ref{sec:base_model} describing our disc model and the implementation of flybys of nearby massive stars, here we now present the results of simulations of disc evolution, exploring the effects of different parameters on the evolutionary tracks of protoplanetary discs. We mainly focus on the parameters surrounding the flybys themselves, i.e. flyby time and the flyby time-scale, as well as the effects of the maximum UV field strength that discs experience.

\begin{figure}
\centering
\includegraphics[scale=0.5]{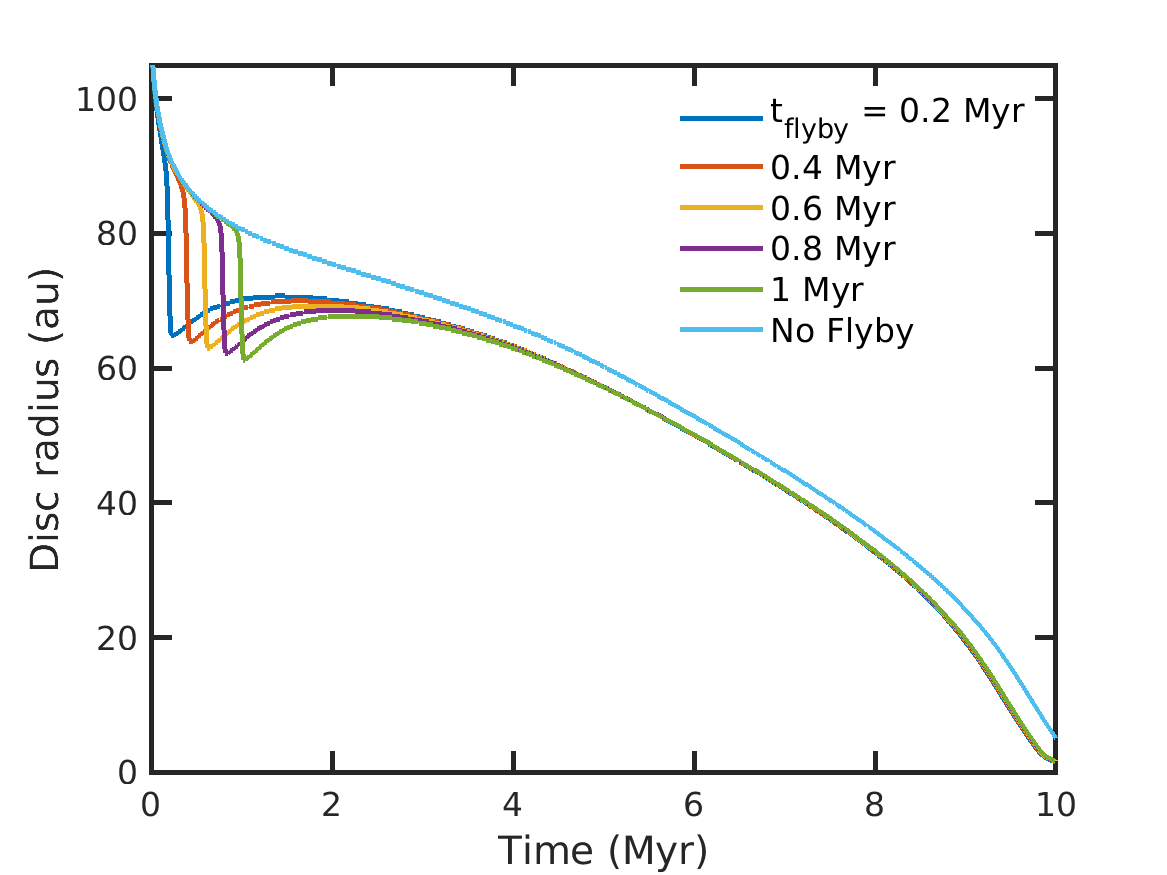}
\caption{Temporal evolution of the disc radius containing 90\% of the mass for discs undergoing flybys at different times in the disc lifetimes. These include: 0.2 Myr (blue), 0.4 Myr (red), 0.6 Myr (yellow), 0.8 Myr (purple) and 1 Myr (green). We also plot the temporal evolution of a disc where no flyby has occurred for comparison (cyan).}
\label{fig:radii_flyby}
\end{figure}

\subsection{Effect of flybys on disc radii}
Typically, the evolution of disc radii involves a gradual decrease as material is lost to consistent photoevaporative winds, and through transport through the disc and then accretion on to the central star \citep[e.g.][]{Coleman24MHD}. An example of this temporal evolution of the disc radius can be seen by the cyan line in Fig. \ref{fig:radii_flyby}, where with the disc evolving in isolation, it is able to gradually decrease in size as mass is lost through photoevaporative winds and accretion on to the central star. The other lines in Fig. \ref{fig:radii_flyby} show the evolution of disc radii for discs that underwent a flyby of a massive star at various times, with a maximum UV field of $10^{3.5} \rm G_0$, and a flyby duration of 10 Kyr. As can be seen by the discs that included a flyby, their disc radii were efficiently truncated by 20--25$\au$. After the flyby had occurred, and the effective UV field returned to the background values, those discs were then able to slowly expand through viscous expansion. However given the low $\alpha$ values here, consistent with the upper bounds of observations \citep[e.g.][]{Ansdell18,Villenave20,Villenave22}, the rebound of the disc radii is only on the order of $\sim10\au$, until the point where the outer edge of the disc reaches an equilibrium between the viscous expansion and the external photoevaporation rate \citep{Coleman22}.
Note that if the mechanism of angular momentum transport in the disc was driven by MHD winds, rather than viscosity, then the rebound noted here would not occur, since MHD winds result in the inward movement of material instead of expansion \citep{Tabone22}. Additionally, if the strength of the viscous parameter $\alpha$ was weaker, e.g. $\le10^{-4}$ then viscous expansion would be significantly reduced, with the discs remaining more compact as they equilibrium point between viscous expansion and the mass lost through photoevaporative processes would occur at smaller disc radii.
After the rebound of the disc radius, those discs then continue to evolve similarly to the disc without a flyby, but at a slightly more evolved state with consistently smaller disc radii, and ultimately, slightly shorter disc lifetimes.

\begin{figure}
\centering
\includegraphics[scale=0.5]{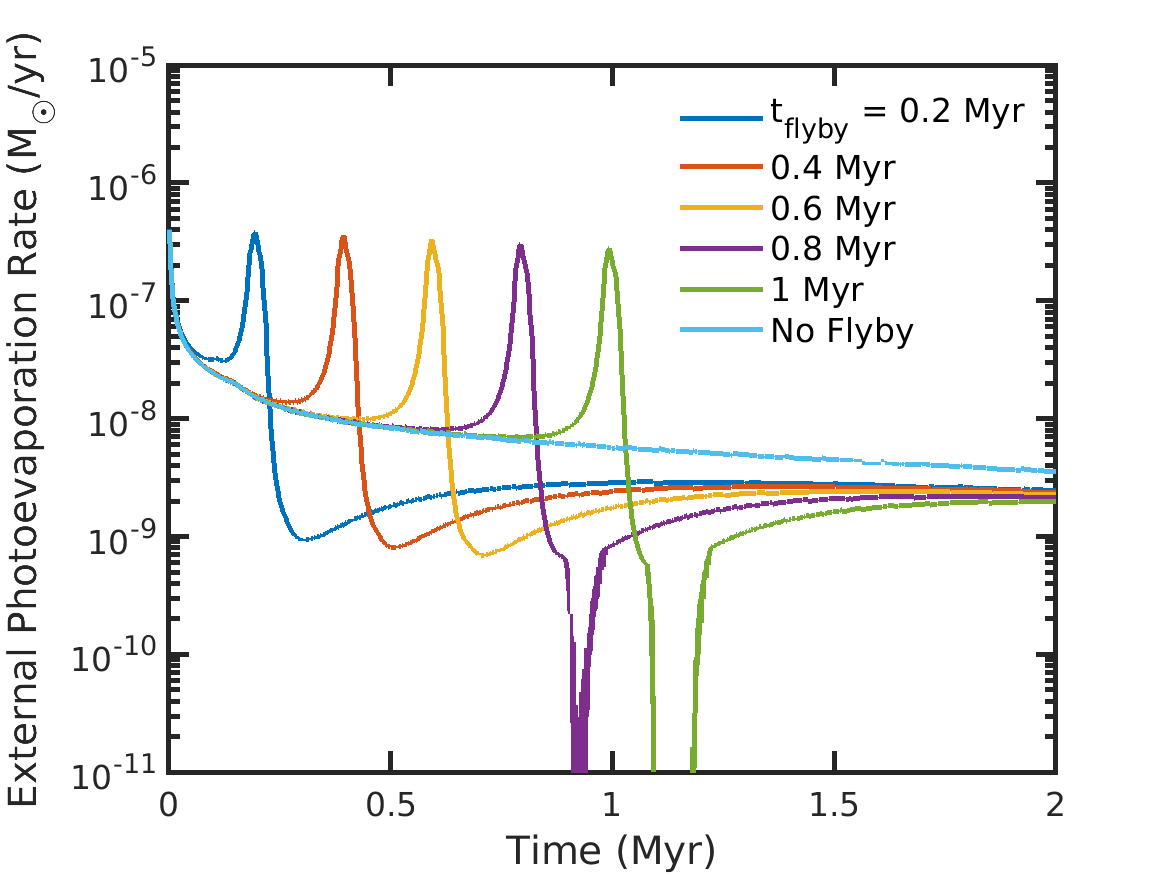}
\caption{External photoevaporation rates for discs undergoing flybys at different times in the disc lifetimes. These include: 0.2 Myr (blue), 0.4 Myr (red), 0.6 Myr (yellow), 0.8 Myr (purple) and 1 Myr (green). We also plot the temporal evolution of a disc where no flyby has occurred for comparison (cyan).}
\label{fig:evap_ext_flyby}
\end{figure}

Whilst Fig. \ref{fig:radii_flyby} showed the evolution of the disc radii for discs undergoing flybys at different times in their lifetimes, Fig. \ref{fig:evap_ext_flyby} shows the external photoevaporation rates for those discs. The cyan line again shows the disc where no flyby occurred and clearly shows the gradual decrease of the external photoevaporation as the disc becomes smaller and less massive. For the other profiles however, the effect of the flybys are clear, where the external photoevaporation rate is seen to increase by at least an order of magnitude around the times of each flyby. Once the flyby star has passed, the external photoevaporation rate drops to a weaker level than before the flyby, and also less than for that where no flyby has occurred. This is the direct result of the discs being significantly truncated, with the UV radiation now unable to liberate more gravitationally bound gas. Additionally, the external photoevaporative wind is also competing with the internal photoevaporative wind in these regions, further reducing their effectiveness. This is especially clear for the purple and green lines, showing flybys at later times, where the external photoevaporation rate drops to negligible levels with internal photoevaporation dominating the mass loss rates at those radii. As the disc rebounds further out due to viscous expansion, the external photoevaporation rates increase to a quasi-steady value before gradually reducing as the discs evolve. They however do not reach the level of that for the disc without a flyby, due to the permanent loss of mass and radius of the disc that came about because of the extreme flyby event.

Interestingly, the effects of the timing of the flybys on both the disc radii and external photoevaporation rate can not be identified, long after each flyby has occurred. This is due to the profiles following similar tracks once the discs have undergone viscous expansion and then begun to be truncated again from the steady external photoevaporation rates.

\begin{figure}
\centering
\includegraphics[scale=0.5]{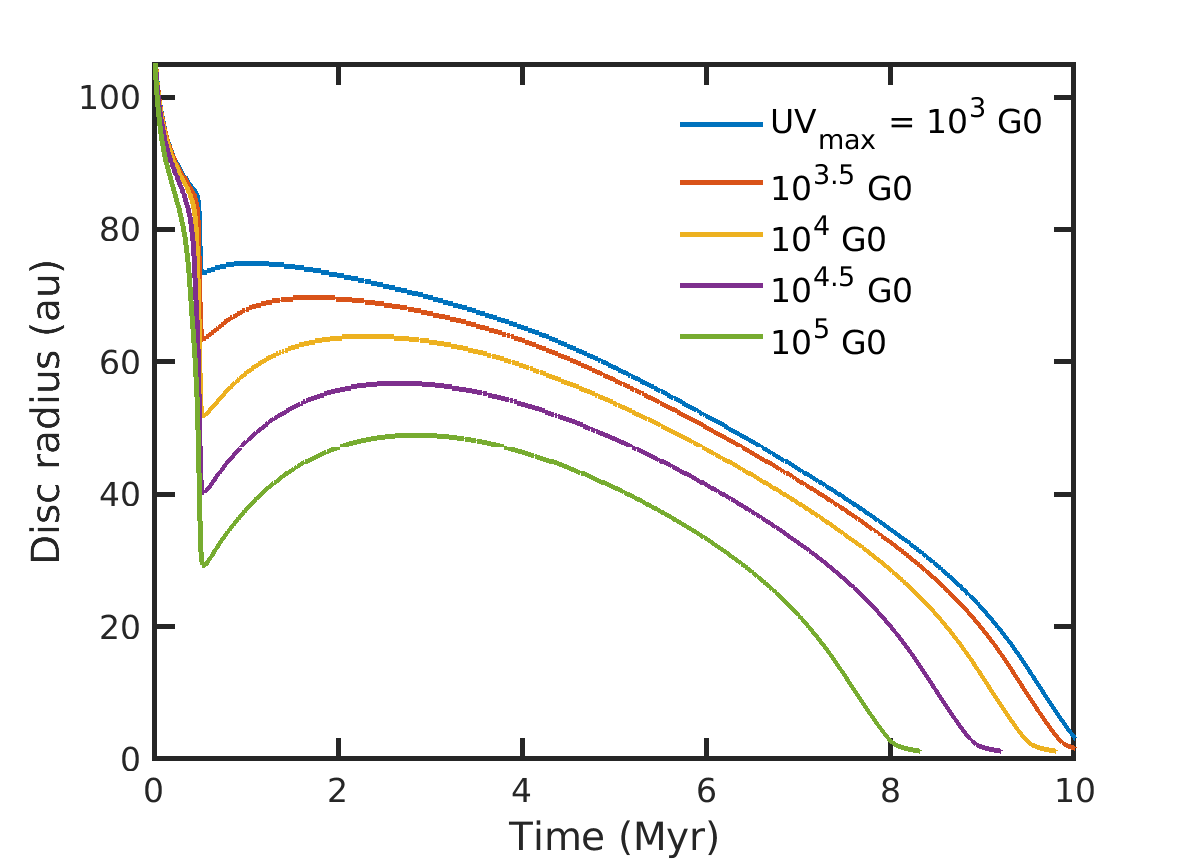}
\caption{Temporal evolution of the disc radius containing 90\% of the mass for discs undergoing flybys of different maximum UV field strengths. These include: $10^{3} \rm G_0$ (blue), $10^{3.5} \rm G_0$ (red), $10^{4} \rm G_0$ (yellow), $10^{4.5} \rm G_0$ (purple) and $10^{5} \rm G_0$ (green).}
\label{fig:radii_uv}
\end{figure}

\begin{figure}
\centering
\includegraphics[scale=0.5]{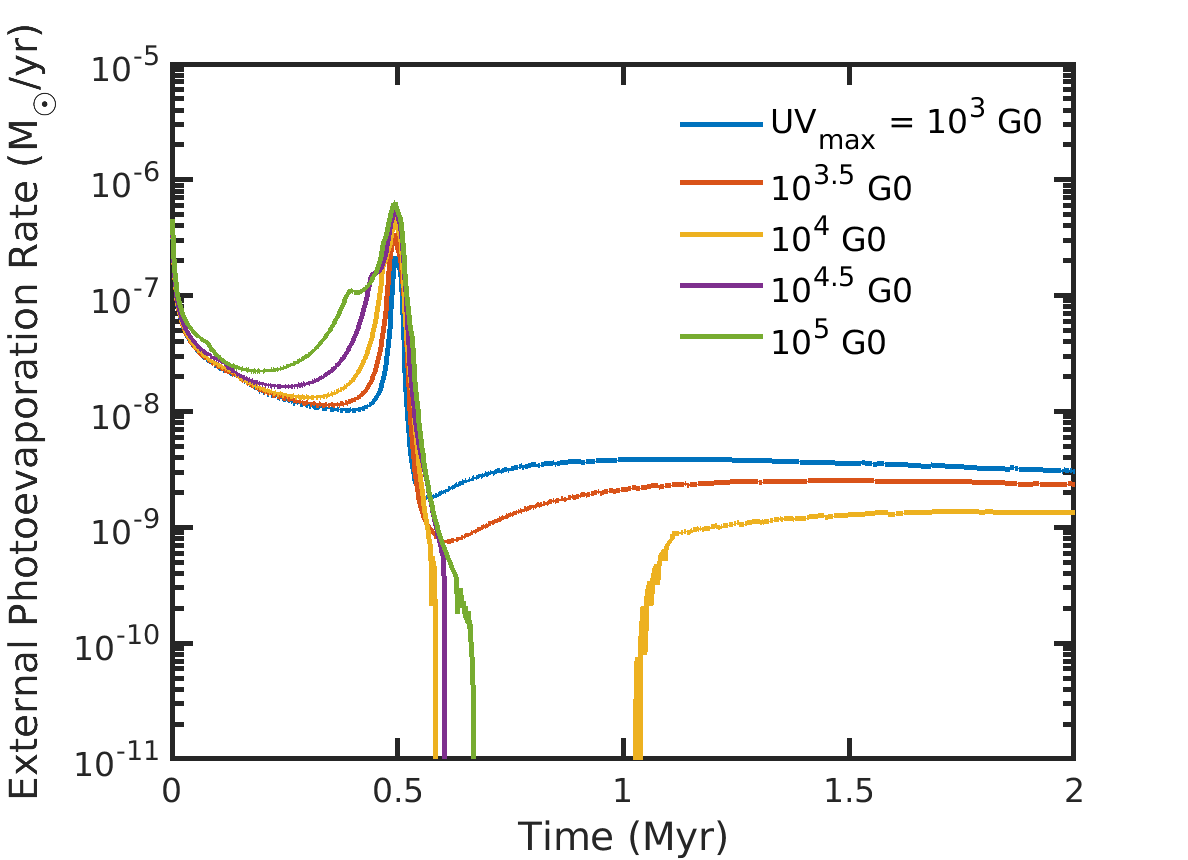}
\caption{External photoevaporation rates for discs undergoing flybys of different maximum UV field strengths. These include: $10^{3} \rm G_0$ (blue), $10^{3.5} \rm G_0$ (red), $10^{4} \rm G_0$ (yellow), $10^{4.5} \rm G_0$ (purple) and $10^{5} \rm G_0$ (green).}
\label{fig:evap_ext_uv}
\end{figure}

\subsection{Role of maximum UV field strength}

We now explore the effects of different maximum FUV strengths from the flyby events on the evolution of the protoplanetary discs. The difference in such strengths can be due to different stars of different masses being the flyby star, or the flyby event occurring at a lesser distance to the simulated protoplanetary disc. Figure \ref{fig:radii_uv} shows the evolution of the disc radius taken at the point that contains $90\%$ of the mass for different maximum FUV strengths. The background UV field was set at 100 $\rm G_0$, the flyby time and duration was set to 0.5 Myr and 10 Kyr respectively. The viscous $\alpha$ value was equal to $10^{-3}$. The different colours show the profiles for varying maximum UV field strengths ranging from $10^3 \rm G_0$ (blue) to $10^5 \rm G_0$ (green). As would be expected, when the maximum UV field strength is increased, the effects on the disc radii are more pronounced as they are truncated to a much larger extent. For example, the disc denoted by the red line where the maximum UV field was set to $10^{3.5} \rm G_0$ was truncated down to $65\au$ from $\sim90\au$, as was also shown in Fig. \ref{fig:radii_flyby}. However, when comparing that to the disc shown by the green line, there the truncation was down to $\sim30\au$ since the maximum UV field was set to $10^5 \rm G_0$. This shows the extreme effect that even a short flyby can have on disc radii, and subsequently disc mass as significant amounts of gas is lost due to external photoevaporative winds.

Interestingly, the extent to which a disc viscously rebounds varies with the strength of the maximum UV field. As was shown in Fig. \ref{fig:radii_flyby}, the flyby time had little effect on the amount of reboundedness for discs undergoing a flyby of strength $10^{3.5} \rm G_0$, also shown by the red line in Fig. \ref{fig:radii_uv}. As can be seen in Fig. \ref{fig:radii_uv} the extent to which a disc rebounds increases as the strength of the maximum UV field increases. For example, the weakest flyby examined here, equal to $10^{3} \rm G_0$ as shown by the blue line, only rebounds by a few $\au$. On the other hand, the disc experiencing the strongest flyby, $10^{5} \rm G_0$ as shown by the green line, was able to rebound from 30$\au$ up to $\sim50\au$. Notice though that the disc was also more substantially truncated, and so the viscous time-scales were much shorter here, that when coupled to the reduced externally driven mass loss rates associated with the weaker background UV field, allowed those discs to expand at a faster rate. However, even though those discs were able to rebound further, the extent of the damage to the disc mass and radius from the intense flyby significantly reduced the lifetime of the disc by $\sim2$ Myr, and also resulted in a large decrease in disc radius compared to if there was a weaker flyby, or no flyby at all. With the permanence of this damage, this introduces added degeneracy into disc population studies when exploring statistics such as disc fractions over time \citep[e.g][]{Coleman22}.

As is shown in Fig. \ref{fig:radii_uv}, the maximum UV strength has a large effect on the evolution of disc radii over time. In Fig. \ref{fig:evap_ext_uv} we show the temporal evolution of the external photoevaporative mass loss rate, with the colours denoting the different UV strengths of the flyby stars. The timing of the flyby at 0.5 Myr is clearly evident with the spike in the mass loss rates, whilst the effect of the maximum UV strength can be seen with the increase at the peak of the flyby. Additionally, as the discs viscously rebound, the external mass loss rates increase to a new base level. This can be seen by the blue and red lines for example. But, as the maximum UV strength increases, the time it takes for the external mass loss rates to reach a nominal level, gets longer, and even for the strongest UV fields studied here, there is actually no external mass loss rates after the flyby events. This is due to the disc radii being much smaller for those discs with stronger flybys, that reduces the external mass loss rate. Additionally, the external photoevaporative mass loss rate is also competing with the internal photoevaporative mass loss rate, and so the lack of an external wind can also be due to the numerical choice here of selecting between the maximum of the two mass loss rates. However, even if there was a contribution to the photoevaporative winds from both internal and external sources, the rate would be much reduced compared to that if there was no flyby.

\begin{figure}
\centering
\includegraphics[scale=0.5]{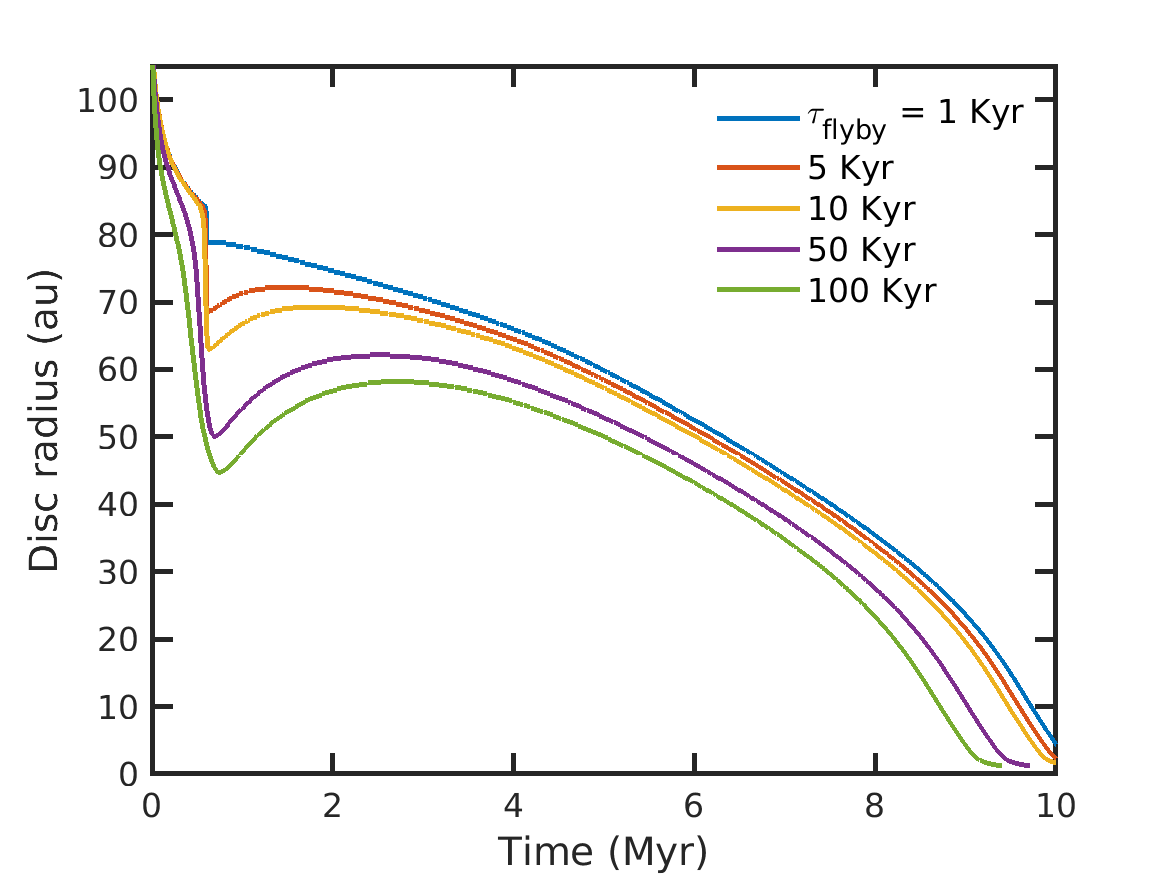}
\caption{Temporal evolution of the disc radius containing 90\% of the mass for discs undergoing flybys of different time-scales. These include: 1 Kyr (blue), 5 Kyr (red), 10 Kyr (yellow), 50 Kyr (purple) and 100 Kyr (green).}
\label{fig:radii_timescale}
\end{figure}

\subsection{Importance of flyby time-scales}

The last important factor in determining the effects of a flyby event on an evolving protoplanetary disc, is the time-scale of the flyby itself. If the time-scale is short, then there is less time for the massive star to impact on the evolution of the disc. But as the time-scale increases, i.e. if the flyby star is moving past the disc with a slower relative velocity, then more radiation is able to be imparted on to the disc, resulting in larger mass loss rates for longer, and therefore more mass should be lost due to the flyby event.

To explore this we varied the flyby time-scale from 1--100 Kyr. Figure \ref{fig:flyby_profile} showed the differences of some of these variations in terms of the change of UV field strength over the course of the flyby event and beyond. It is clear there that the total amount of UV radiation is significantly increased for the longer duration flybys. In Fig. \ref{fig:radii_timescale} we again plot the effects on the disc radii of different duration flybys including: 1 Kyr (blue), 5 Kyr (red), 10 Kyr (yellow), 50 Kyr (purple), and 100 Kyr (green). For all of these discs, the flyby event took place after 0.5 Myr, whilst the UV field strength increased from a base of $100 \rm G_0$ up to $10^{3.5} \rm G_0$. It is clear that as the flyby duration increased, the effects on the discs became more profound, with the longer time-scales significantly truncating the disc down to $\sim 50 \au$. This shows the importance of the flyby time-scale, as well as that of the maximum UV field strength of the flyby star and the distance to it, in determining the evolution of the protoplanetary discs. Indeed, looking at the final evolution of those discs in Fig. \ref{fig:radii_timescale}, the discs that had longer duration flyby events ultimately dispersed faster, with final lifetimes being decreased by $\sim 1$ Myr. This could possibly add further degeneracy to disc evolution studies involving disc lifetimes and disc fractions \citep[e.g.][]{Coleman22}, as it shows that the dynamical history of the stars and discs in stellar clusters may be important if flyby events are frequently occurring, and if sufficient UV radiation is imparted on to evolving protoplanetary discs.

\begin{figure*}
\centering
\includegraphics[scale=0.45]{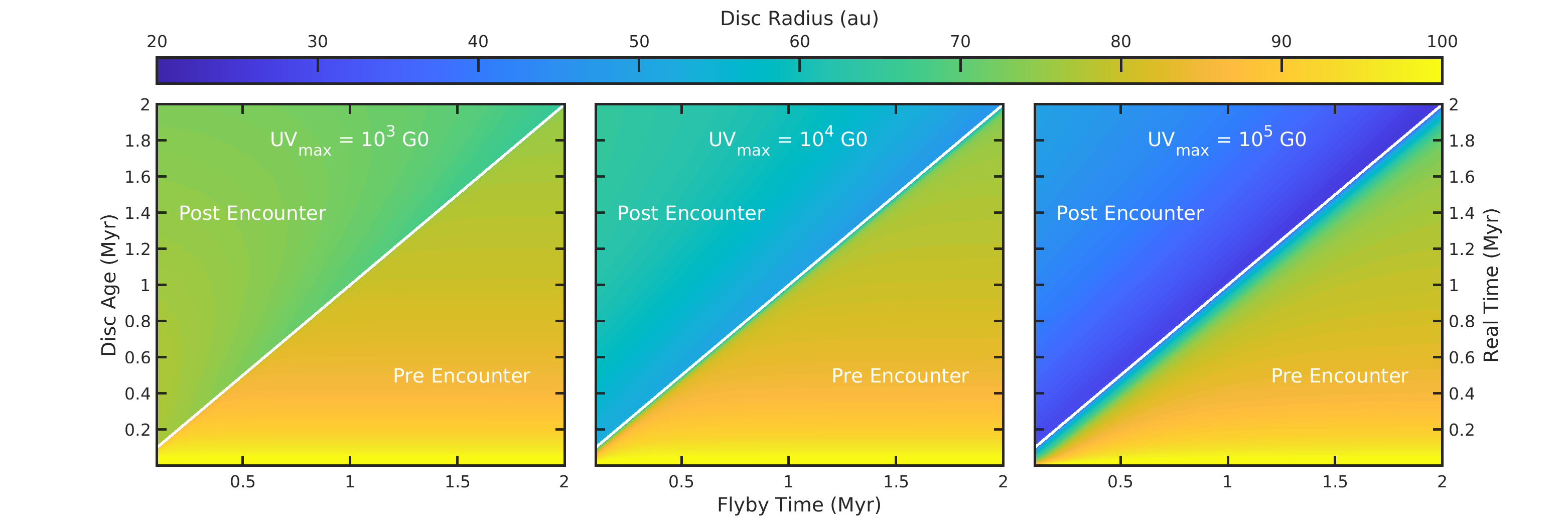}
\caption{Contour plots showing disc radii as a function of the flyby time and the disc age. Different panels show the disc radii for different maximum UV field strength with $10^3\rm G_0$ shown in the left-hand panel, $10^4\rm G_0$ in the middle panel, and $10^5\rm G_0$ in the right-hand panel. The white diagonal line in each panel denotes when the flyby time is equal to the disc age, i.e. that is the time for that disc that the flyby is occurring. Regions above the white line show where discs have already had an encounter in their lifetime, whilst those below who discs where a flyby event has yet to occur.}
\label{fig:radii_contours}
\end{figure*}

\section{Observable features in stellar clusters}
\label{sec:observables}

The section above explored the effects on individual protoplanetary discs of a flyby from a massive star. We now extrapolate those results to explore the effects on disc properties across a stellar cluster. As a star moves through a cluster, it will interact with different protoplanetary discs at different times and distances, akin to the different flyby times shown in Fig. \ref{fig:radii_flyby} for example. With the variance in the interaction times, and the significant effects on disc radii of a flyby, as was shown in Sect. \ref{sec:results}, then there should be distributions of disc radii as a star passes through a cluster.

Figure \ref{fig:radii_contours} shows contour plots of disc radii as a function of the flyby time ($x$-axis) and the disc age ($y$-axis). The different panels denote different maximum UV strengths of the flyby star, including $10^3 \rm G_0$ (left-hand panel), $10^4 \rm G_0$ (middle panel), and $10^5 \rm G_0$ (right-hand panel). The white diagonal line in each panel denotes when the flyby time is equal to the disc age, i.e. that is the time for that disc that the flyby is occurring. For reference, those points below the white line, are classed as being ``pre-encounter'', i.e. the flyby is yet to happen, whilst those above the line are classed as ``post-encounter'', i.e. the flyby has already occurred. The flyby time-scale was set to 10 Kyr.

Looking initially at the left-hand panel of Fig. \ref{fig:radii_contours}, showing the evolution of disc radii as discs undergo a $10^3 \rm G_0$ flyby event, the effect of the flyby event on the disc radii is clear as the colour of the contour is different above and below the white line. The evolution of the discs can also be seen as they age, with the radii decreasing from 100 $\au$ at the start of their lifetime, down to 80$\au$ after 2 Myr, for those without a flyby within that time. For those that experienced a flyby, their disc radii can clearly be seen to  be dropping to $\sim 60 \au$. The slight viscous rebound of the disc can also be seen as those that have an early encounter are able to slowly grow. As the intensity of the flybys increase when going from the left-hand panel to the middle and then the right-hand panel, the change in colour above and below the white line dramatically increases. For the left-hand panel, the discs above the white line are generally of size $\sim$60--$70\au$, green in colour, whilst those for the right-hand panel are more significantly truncated at between 20--40$\au$, as shown by the blue contours. When comparing to those discs below the white line, this shows the extreme effects that a flyby of an energetic star has on protoplanetary discs. Additionally, for the strongest flybys in the right-hand panel, the effects of the flyby can be seen before the peak flyby time, with the discs already truncated by $\sim20\au$ just below the white line.

\subsection{Variation in disc radii either side of a moving flyby star}
\begin{figure}
\centering
\includegraphics[scale=0.5]{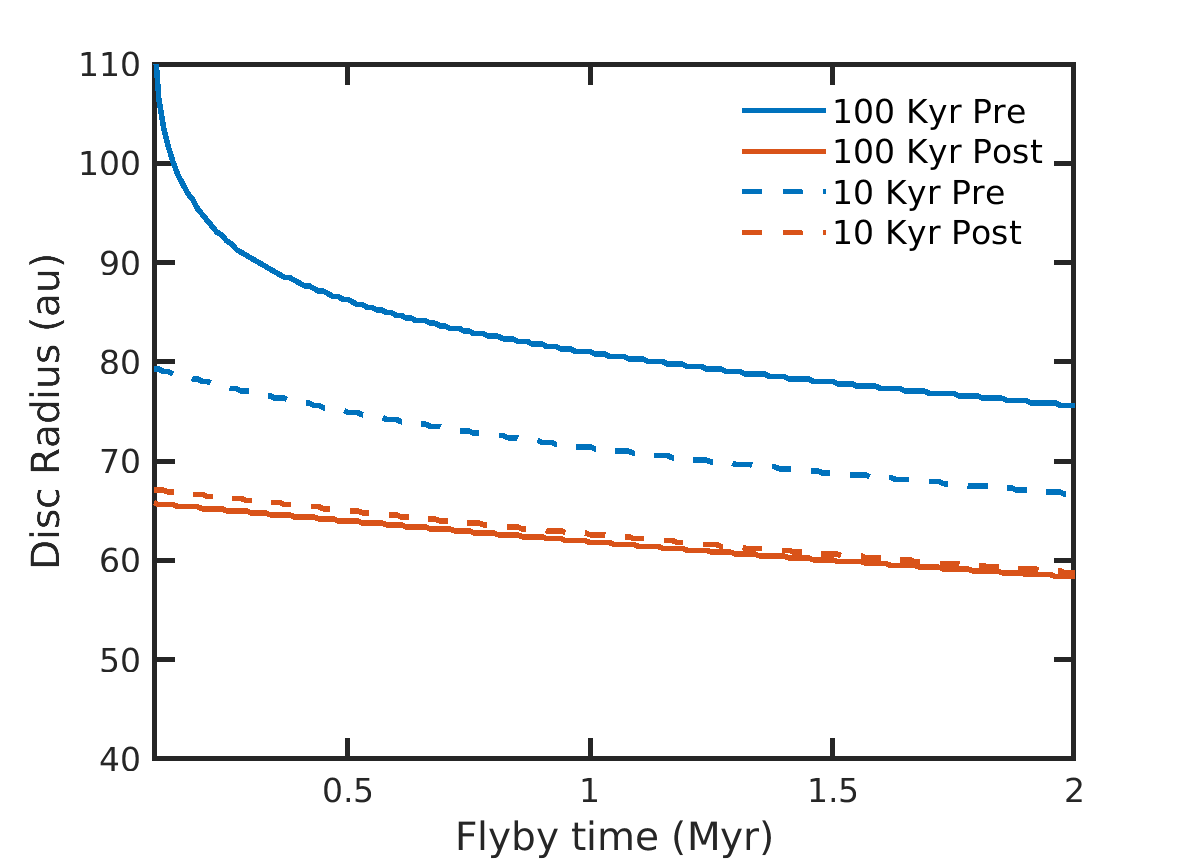}
\caption{Disc radii as a function of the flyby time, taken 100 Kyr (solid lines) and 10 Kyr (dashed lines), before (blue lines) and after (red lines) flyby encounter times.}
\label{fig:radii_comparison}
\end{figure}

With Fig. \ref{fig:radii_contours} showing significant differences in disc radii pre- and post-encounter of a flyby star, then there should be differences in disc radii as a star moves through a cluster. Those stars in front of the star, that have not yet had their disc radii truncated should be substantially larger than those discs behind the star that have had their discs truncated. Figure \ref{fig:radii_comparison} shows the disc radius as a function of the flyby time in the disc's lifetime, taken 100 Kyr (solid lines) and 10 Kyr (dashed lines) before (blue lines) and after (red lines) the flyby event. The flyby time-scale was set to 10 Kyr, and the maximum UV field strength was $10^{3.5} \rm G_0$. As can be seen when comparing the solid lines, 100 Kyr before and after the flyby events, there is typically at least a 20$\au$ difference, highlighting the effects of the flyby on disc radii. The exception to this is early in the disc lifetime, where the discs are still sufficiently extended, and external photoevaporation from the background field has not yet been able to truncate the discs down to an equilibrium with the viscous expansion rate. Interestingly, the early effects of the flyby events can also be seen by the dashed lines, that compare disc radii 10 Kyr before and after the flyby. They show that the discs are significantly truncated even before the peak of the flyby has occurred, as the UV field ramps up from the background field to the maximum UV field from the flyby star. This was also seen to a larger effect for the $10^5\rm G_0$ case (right-hand panel) in Fig. \ref{fig:radii_contours} where the effects of the flybys could be seen below the white line before the peak of the flyby events had occurred.

\begin{figure}
\centering
\includegraphics[scale=0.5]{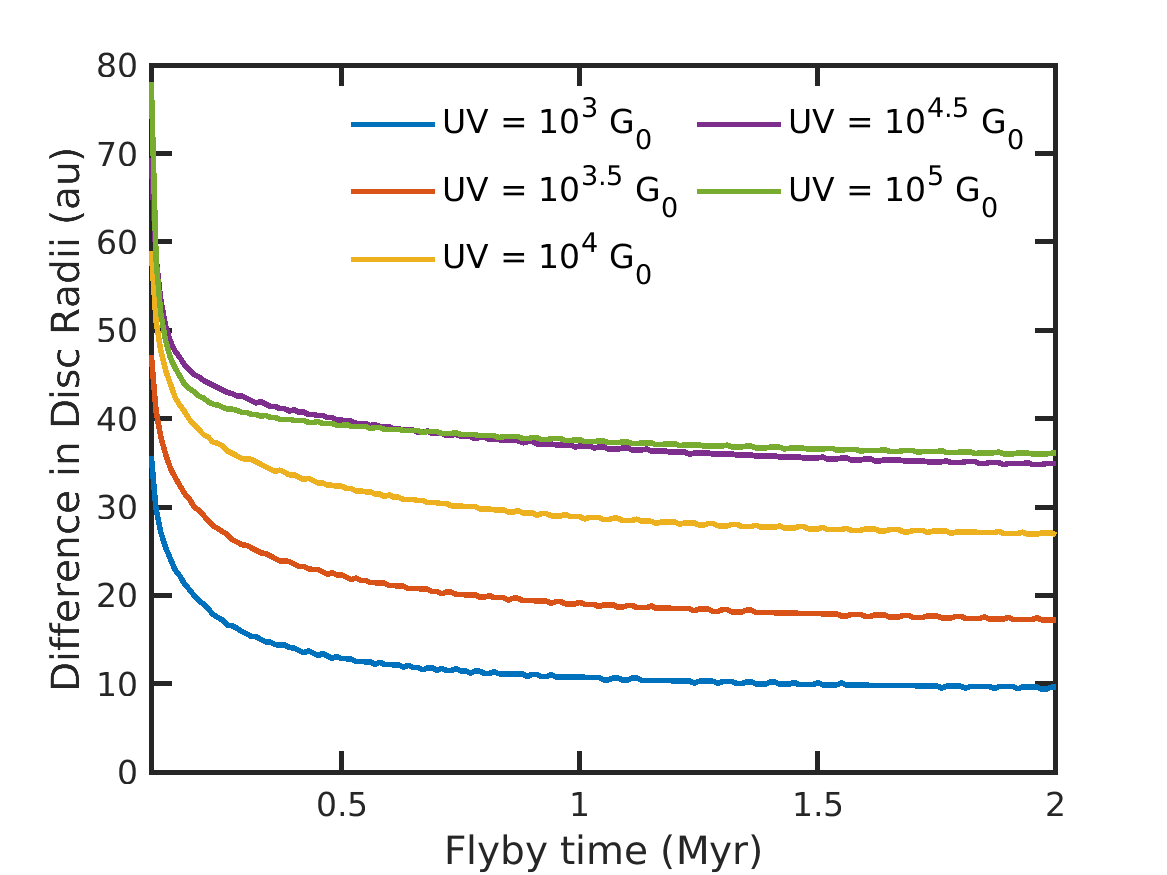}
\caption{Differences between disc radii taken 100 Kyr before and after flyby encounters as a function of the flyby time. Different colours denote different maximum UV field strengths, including: $10^{3} \rm G_0$ (blue), $10^{3.5} \rm G_0$ (red), $10^{4} \rm G_0$ (yellow), $10^{4.5} \rm G_0$ (purple) and $10^{5} \rm G_0$ (green).}
\label{fig:radii_comparison_uv}
\end{figure}

To further show the variation in disc radii, Fig. \ref{fig:radii_comparison_uv} shows the difference between the disc radii 100 Kyr before and after a flyby event as a function of the flyby time. The different colours represent different maximum UV field strengths ranging from $10^3 \rm G_0$ (blue lines) to $10^5 \rm G_0$ (green lines), with the red lines showing the differences in disc radii for those found in Fig. \ref{fig:radii_comparison}. Again, the flyby time-scale was set to 10 Kyr. As can be seen by the red line, apart from when the disc is very young, the difference in disc radii 100 Kyr before to that after the flyby events are typically around 20$\au$. As discussed above, this is for a maximum UV field strength of $10^{3.5} \rm G_0$. When increasing the maximum strength, the difference in disc radii are found to also increase, and thus for a $10^5 \rm G_0$ flyby, the differences are now at $\sim 40 \au$. Even for the weakest flyby case here of $10^3 \rm G_0$, the difference in disc radii before/after a flyby event was at least 10 $\au$. With such large differences in disc radii, observations of discs in the vicinity of a massive star, should yield distributions in disc radii respective to the massive stars' direction of travel.

Our discussion so far has focused on the gas disc, but we note that the dust disc is also expected to be truncated by external photoevaporation \citep{Sellek20}, though this can be affected by the presence of pressure bumps \citep{Garate2024}. High resolution ALMA continuum surveys of disks where a runaway OB star is suspected would therefore provide an interesting test of the expected variation of disk radii throughout the region while simultaneously controlling for disk substructures, which would be of interest in their own right.

\subsection{Interacting with different age groups of stars}

Since there is ongoing star formation in stellar clusters, where this could occur in bursts that form groups of stars, there could also be the possibility that as a massive stars moves through a cluster, it encounters groups of stars of different ages at different times. For example, it could move through a young group of stars, that have only just formed, before then immediately passing through a group of older stars. Since the older stars will have had time to evolve, then the effects of the flyby star on their disc properties will be different to what they would be for younger stars, as is shown in Figs. \ref{fig:radii_contours} and \ref{fig:radii_comparison}, where the older discs undergo increased truncation compared to the younger discs. As we noted above, there will be a difference in disc radii in front of to behind a flyby star, and so then the age and evolution of those stars or group of stars may be important.

\begin{figure}
\centering
\includegraphics[scale=0.5]{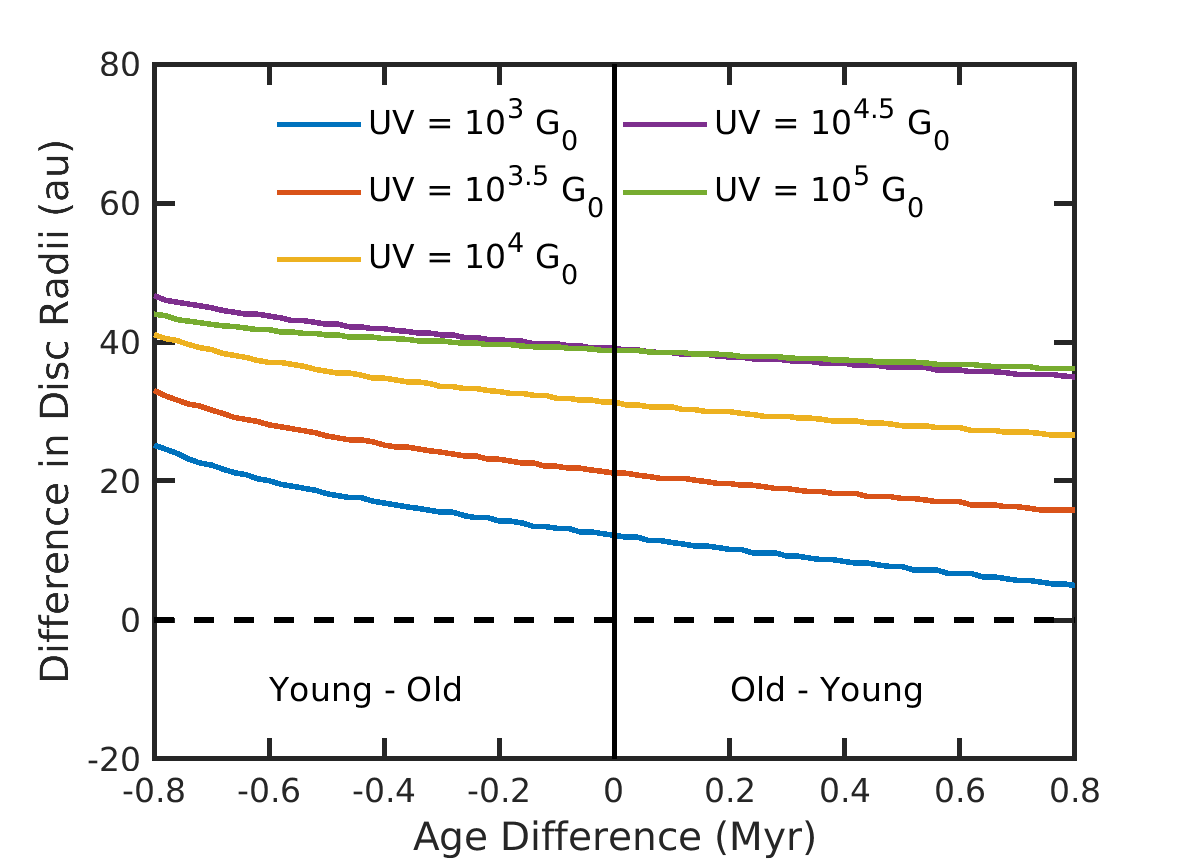}
\caption{Differences between disc radii taken 100 Kyr before and after flyby encounters as a function of the age difference between stars in front of to those behind a massive star moving through a cluster. The average flyby time was set 0.6 Myr for all examples. Different colours denote different maximum UV field strengths, including: $10^{3} \rm G_0$ (blue), $10^{3.5} \rm G_0$ (red), $10^{4} \rm G_0$ (yellow), $10^{4.5} \rm G_0$ (purple) and $10^{5} \rm G_0$ (green).}
\label{fig:ages_comparison}
\end{figure}

In Fig. \ref{fig:radii_comparison_uv} we showed the differences in disc radii either side of a flyby star at different flyby times, as a function of the maximum UV field. We assumed the evolution of all of the stars began at the same time, i.e. they are of the same age. Figure \ref{fig:ages_comparison} again shows the differences in disc radii either side of a flyby star for different UV fields, but we now instead plot them as a function of the difference in stellar ages between those stars in front of and behind the flyby star. The flyby time-scale was set to 10 Kyr, whilst the average flyby time was set to 0.6 Myr. With the difference in disc radii being taken for discs 100 Kyr in front of and behind the flyby star, an age difference of 0 would mean that those in front of the star would be 0.6 Myr old and would encounter the flyby star after 0.7 Myr, whilst those behind the flyby star would have interacted with it after 0.5 Myr. For those discs where the age difference would be -0.8 Myr, then the discs in front of the flyby star would be 0.2 Myr old, whilst those behind the star would be 1 Myr. This is denoted by the \textit{Young - Old} label in Fig. \ref{fig:ages_comparison}. The opposite would be the case for positive age differences.

As can be seen on the left side of Fig. \ref{fig:ages_comparison} where there are young stars in front of the flyby star, and old stars behind it, the differences in disc radii extend from $\sim$25--45$\au$ depending on the maximum UV field strength of the flyby star. For the opposite case, where older stars are in front of the flyby star and young stars are behind it, then the difference in radii falls to being between $\sim$5--35$\au$, again with the variance coming from the maximum UV field strength of the flyby star. In both cases, the smaller differences correspond to weaker stars, whilst the larger differences arise due to stronger interactions that effectively strip the discs of material. For the weaker interactions, with the maximum UV strength of $10^3 \rm G_0$, this is where there is the largest difference in disc radii, ranging from 25$\au$ for an age difference of -0.8 Myr to only 5$\au$ for an age difference of 0.8 Myr. This highlights the importance of knowing the age of the stars when comparing disc sizes and other properties in stellar clusters. Interestingly, for more extreme flyby events, i.e. the green line showing $10^5 \rm G_0$, there isn't much difference between disc radii with young/old stars in front of/behind the flyby star. This is due to the extreme nature of the event where the increase in the radiation effectively truncates the disc long before the peak flyby event, reducing the disc radii of all discs down to a smaller size. After the event, the disc radii are effectively truncated to where external mass loss rates drop due to the discs being so truncated, that most of the gas remains bound to the central star, and it becomes hard for the radiation to launch a wind.

Ultimately, Fig. \ref{fig:ages_comparison} shows that the ages of the population of stars interacting with a flyby star has an effect on understanding the disc radii distributions across a cluster. Additionally it shows that there is a degeneracy between the age difference between stars, and the strength of the maximum UV environment, when using disc radii to explore their evolution within the cluster.

\subsection{Effects of other parameters}
It is not only the differences in the external UV field strength or the variance in ages within a cluster that can affect the differences in disc radii in front of and behind a massive star moving through a cluster. Indeed, the stellar mass and the initial compactness of the disc affect how protoplanetary discs evolve, with less massive stars being especially vulnerable to strong levels of external photoevaporation \citep{Coleman22}.

\begin{figure}
\centering
\includegraphics[scale=0.5]{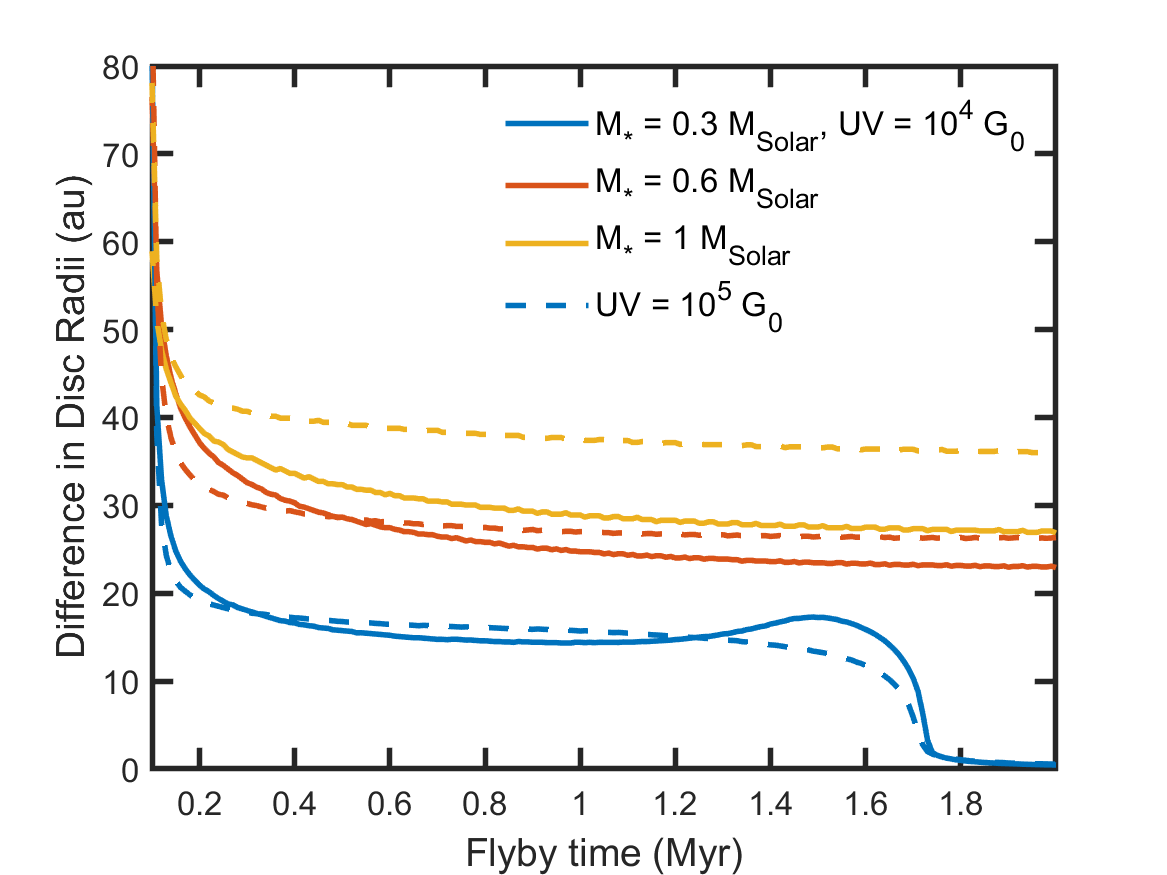}
\caption{Differences between disc radii taken 100 Kyr before and after flyby encounters as a function of the flyby time. Different colours denote different stellar masses, including: $0.3 \msun$ (blue), $0.6 \msun$ (red), and $1 \msun$ (yellow). Solid lines denote discs evolving in environments with a UV field strength of $10^4 \rm G_0$, with dashed discs denoting UV field strengths of $10^5 \rm G_0$.}
\label{fig:radii_comparison_stellar_mass}
\end{figure}

To explore the effects of stellar mass, we perform an identical suite of simulations to that presented above, but use stellar masses of 0.3 and 0.6$\msun$. Similar to Solar mass stars, stars of lower mass also undergo significant truncation as a massive star passes by. The extent of the truncation is also found to be increased for stars that have just experienced a flyby, compared to those awaiting a flyby. In fig. \ref{fig:radii_comparison_stellar_mass}, we show the differences in disc radii either side of a flyby star at different flyby times, with different colours representing different stellar masses. Solid lines show flybys with maximum UV field strengths of $10^4 \rm G_0$, whilst dashed lines show for UV field strengths of $10^5 \rm G_0$. It is clear that there are still differences in disc radii in front of compared to behind a flyby star for all stellar masses, even at flyby times of up to 1.5 Myr. As the stellar mass decreases, the extent of the differences also decreases, since discs around lower mass stars are already significantly truncated from the weaker ambient UV field. This earlier truncation of the discs results in there being little difference for discs in front of and behind at the bottom right of fig. \ref{fig:radii_comparison_stellar_mass} for 0.3 $\msun$ stars after 1.7 Myr. Nonetheless, the role of stellar mass provides a variance on the difference in disc radii of $\sim 20\au$ highlighting the importance of knowing the stellar mass of observed discs, when using flybys of massive stars to infer how protoplanetary discs evolve.

\begin{figure}
\centering
\includegraphics[scale=0.5]{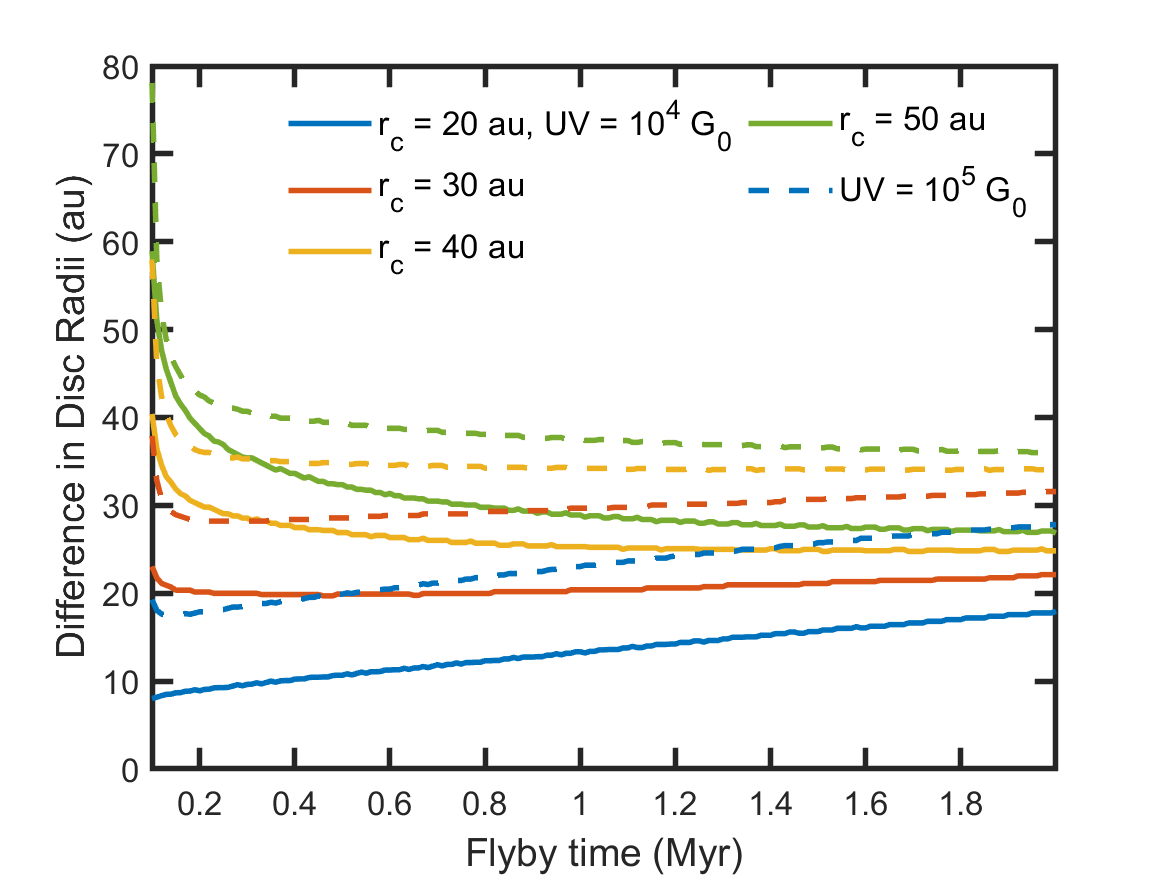}
\caption{Differences between disc radii taken 100 Kyr before and after flyby encounters as a function of the flyby time. Different colours denote different initial scale radii representing the initial compactness of the discs, including: $20 \au$ (blue), $30 \au$ (red), $40 \au$ (yellow), and $50 \au$ (green). Solid lines denote discs evolving in environments with a UV field strength of $10^4 \rm G_0$, with dashed discs denoting UV field strengths of $10^5 \rm G_0$.}
\label{fig:radii_comparison_rc}
\end{figure}

As well as the stellar mass, the initial compactness of the disc defined by the scale radius, affects the level of truncation of the disc as a massive star performs a flyby. This is due to more initially compact discs having a larger percentage of their mass budget deeper in the gravitational well, more protected from the effects of external photoevaporation. Nonetheless, there are still differences in the disc radii in front of and behind a massive star as it moves through a cluster. To explore the effects of this compactness, we also ran a suite of simulations using the same discs as those simulated above, but the initial scale radius was set to either 20, 30, or 40 $\au$. Figure \ref{fig:radii_comparison_rc} shows the differences in disc radii either side of a flyby star at different flyby times, with different colours representing different initial scale radii of the discs, ranging from 20 $\au$ (blue) to 50 $\au$ (green). Solid lines show flybys with maximum UV field strengths of $10^4 \rm G_0$, whilst dashed lines show for UV field strengths of $10^5 \rm G_0$. As can be seen in Fig. \ref{fig:radii_comparison_rc} the differences in disc radii in front of and behind a flyby star decreases as the discs are initially more compact, a result of the mass being more gravitationally bound. Nonetheless, there are still differences of more than $20 \au$ for disc with $r_{\rm c} \ge 30\au$. Even those discs that had an initial compactness of $r_{\rm c}=20\au$ showed differences of at least 10 $\au$. With the variance in disc radii again being between 10--40$\au$ for maximum UV field strengths of $\ge 10^4 \rm G_0$, similar to the variance seen in Fig. \ref{fig:radii_comparison_stellar_mass} for different stellar masses, there will be considerable degeneracy when resolving which parameters are determining the differences in disc radii in front of and behind a flyby star. Should some of these parameters be observationally calculated, i.e. stellar mass or age, that would break some of the degeneracy and the differences in disc radii could inform on other properties of protoplanetary disc evolution, such as the initial compactness of the disc.

\section{Specific Example - 42 Ori and NGC 1977}
\label{sec:42ori}

The sections above explored the effects of a flyby event for theoretical protoplanetary discs in stellar clusters. We now examine a specific observed cluster as an example of where the effects of flybys of massive stars may have an impact on protoplanetary disc evolution. In order to calculate the evolution of those discs, it is important to understand the movement of stars within the cluster. Recent observations of NGC 1977, and in particular the young stars in the vicinity of 42 Ori, have determined their proper motions, and thus have enabled their trajectories within the cluster to be known (Kim et al. in prep). Note, however that this is with the assumption that all of the stars are at similar distances from the Solar system, and that their 2D projected distances are good estimations for their 3D distances as predicted by \citet{Anania25}.

\begin{figure}
\centering
\includegraphics[scale=0.55]{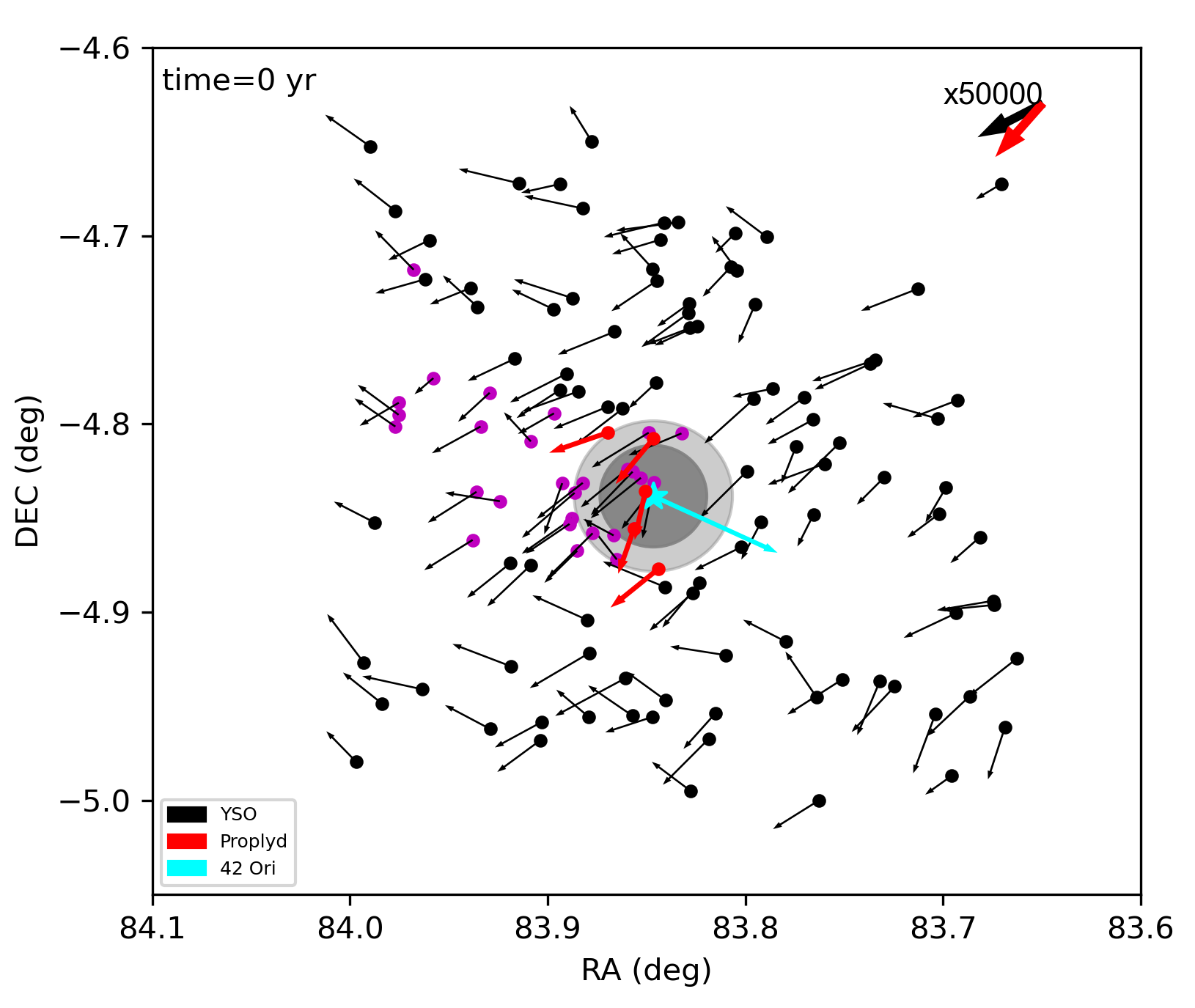}
\caption{Proper motion of young stellar objects (black) and proplyds (red) from Kim et al. (in prep) from Gaia DR3. 42 Ori (B1V) is marked with star symbol in cyan color. Filled circles in magenta color are young stars that have been exposed to G$_{0}\geq1000$ calculated using projected distance from 42 Ori and each source assuming the same proper motion for past $10^5$ years. Red and black arrows on the upper right corner of the plot show median proper motions of the young stellar objects and proplyds. All proper motion vectors are enlarged by 5$\times10^4$.} 
\label{fig:NGC1977_PM}
\end{figure}

NGC 1977 hosts eight known photoevaporating protoplanetary disks (proplyd) \citep{Kim2016} in the vicinity of 42 Ori, a B star. Proper motion of 42 Ori from Gaia DR3 suggests a different travel path compared to those of the young stars, suggesting that it has been moving into the center of young stars in NGC 1977. NGC 1977 is located at the Orion A cloud about 0.5$^\circ$ north of the Orion Nebula Cluster (ONC). Young stars in NGC 1977 are about 1-2 Myr old  (Kim et al. in prep), similar to those in ONC at a similar distance of $\sim$400~pc.
The B1V star, 42 Ori, is the main ionizing source in NGC 1977, providing the FUV field of $\sim10^3\rm G_0$ on average to the proplyds, which is 10--100 times weaker than in the vicinity of $\theta^1$Ori~C (O6.5V) in ONC.
Kim et al. (in prep) used Gaia DR3 and SDSS IV/APOGEE-2 data to study kinematics of young stars in NGC 1977, and found that the mean proper motion of $\sim$100 young stellar objects (including proplyds) in the center of NGC 1977 have been moving toward the south east direction, while 42 Ori has been moving into the central region of NGC 1977 in almost the opposite direction. Figure~\ref{fig:NGC1977_PM} shows the current location of 42 Ori, proplyds, and young stars in the central region of NGC 1977. The purple points show the locations of young stars that had encountered 42 Ori and FUV field $F_{\rm{UV}}\gtrsim10^3 \rm G_0$. The shaded circles show $F_{\rm{UV}}=10^4, 10^3$, and $10^2 \rm G_0$. 

\begin{figure}
\centering
\includegraphics[scale=0.5]{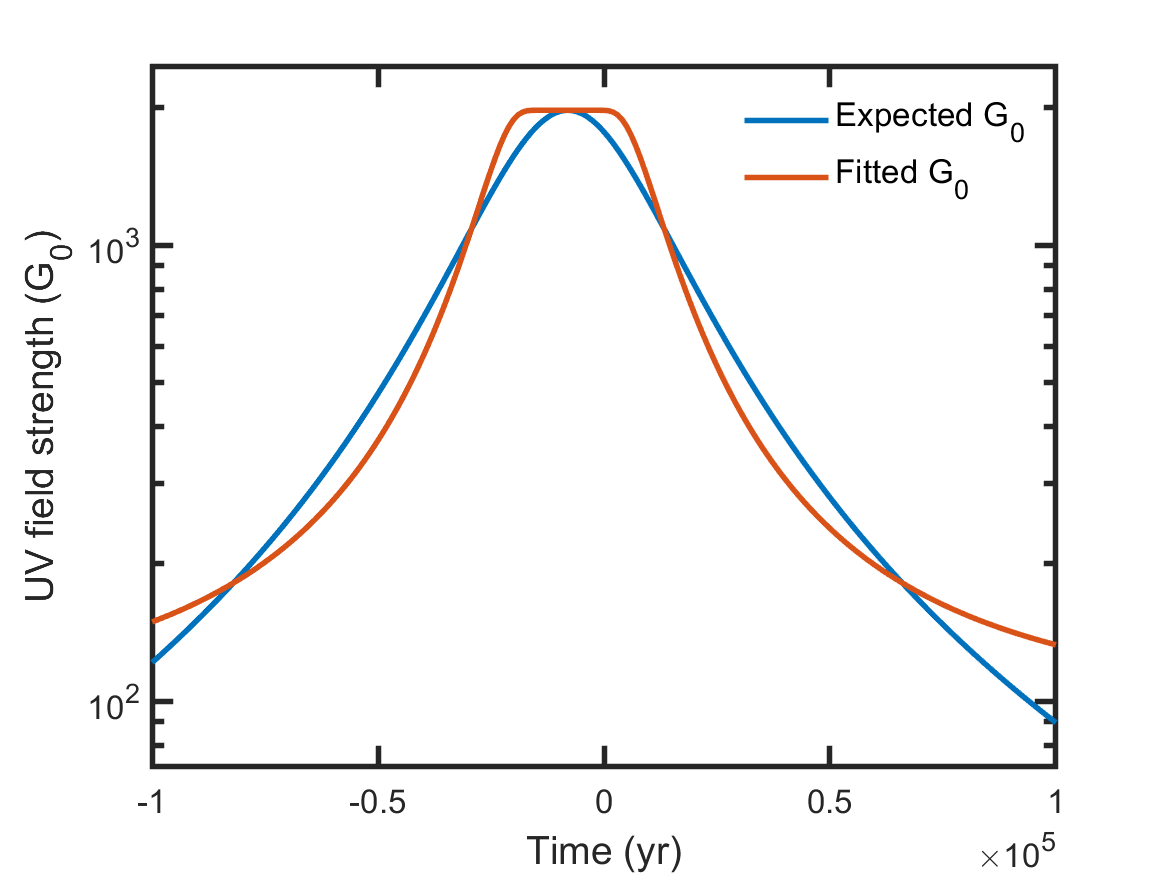}
\caption{Temporal evolution of UV field strength for a specific star in the NGC 1977 Cluster. The blue line shows the expected track based on proper motions in the cluster (Kim et al in prep), whilst the red line shows the best fit to the track using Eq. \ref{eq:uv_implementation}.}
\label{fig:42_ori_example_star}
\end{figure}

By knowing the proper motions of the young stellar objects in NGC 1977, Kim et al. (in prep) determined the dynamical evolution of those stars for the past and future 100 Kyr, and calculated the expected UV radiation field which that star and accompanying disc would experience. An example of this is shown by the blue in Fig. \ref{fig:42_ori_example_star} where UV field gradually increases as the star gets close to the massive star in the region (42 Ori) before then dropping back to a base level as the star moves away from the massive star. By knowing the evolution of the UV field, we were then able to fit the parameters required for Eq. \ref{eq:uv_implementation} so that we could explore the evolution of the protoplanetary disc across it's entire lifetime. This is shown by the red line in Fig. \ref{fig:42_ori_example_star} where the fit was determined for the star represented by the blue line. As can be seen by comparing the red and blue lines in Fig. \ref{fig:42_ori_example_star}, there is a good fit between the calculated and the fitted UV field strengths.

\begin{figure}
\centering
\includegraphics[scale=0.5]{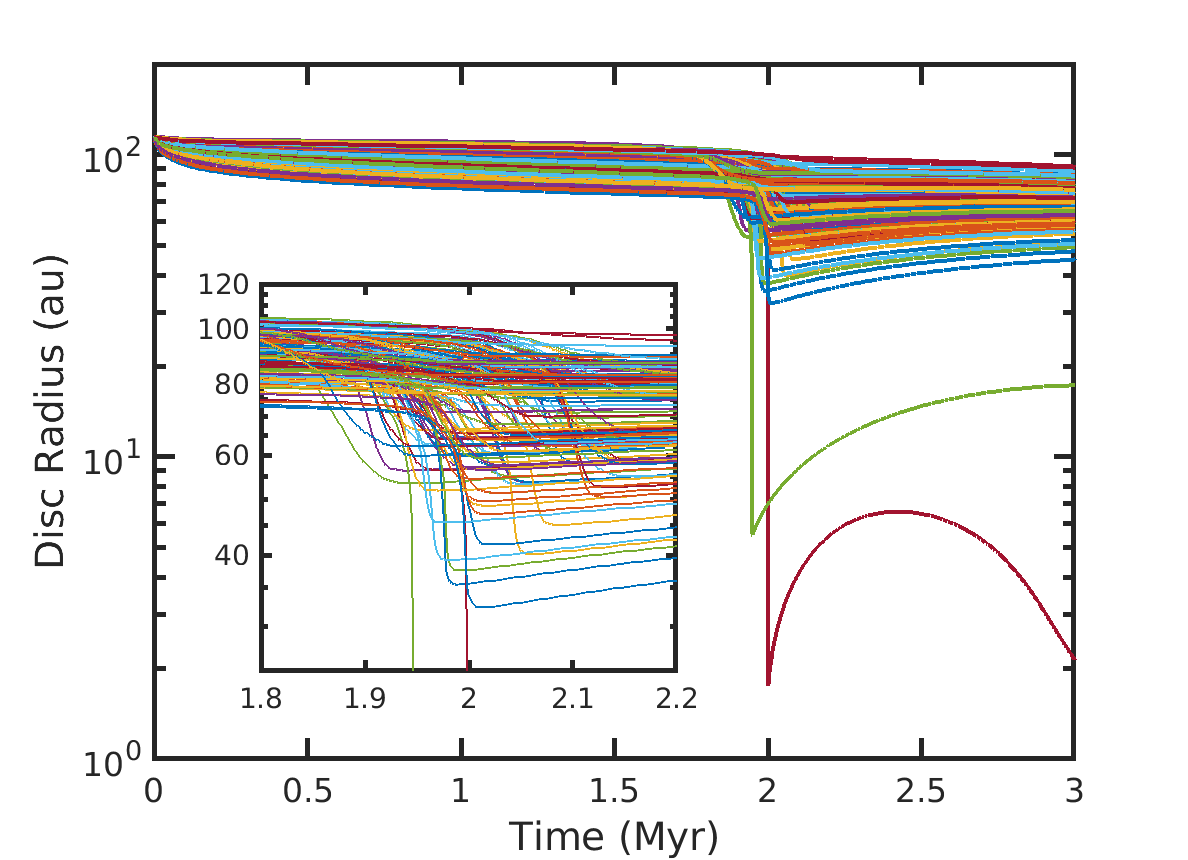}
\caption{Evolution of disc radii over time for the simulated discs around the stars in the sample from NGC 1977.}
\label{fig:42_ori_radii}
\end{figure}

After calculating the UV tracks for each star in our sample from NGC 1977, we then calculated the evolution of their discs, whilst undergoing a flyby event from 42 Ori. With the age of the cluster estimated to be 2 Myr, we assume that all of the flyby events, i.e. like that shown in Fig. \ref{fig:42_ori_example_star}, occur between 1.9--2.1 Myr. Prior to this, the discs are assumed to have evolved in a weak UV environment. Figure \ref{fig:42_ori_radii} shows the evolution of disc radii for the first 3 Myr, where for the first 1.8 Myr, the discs are slowly evolving depending on their background UV field strength. With the background UV field strengths being determined by the fits to Eq. \ref{eq:uv_implementation} (typically between 10--100 $\rm G_0$), this provides some natural variability in the evolution of disc radii, where after 1.8 Myr, the disc radii are all between 70--100 $\au$. As some of the discs are approached by 42 Ori, they are quickly truncated as is shown by the sharp drop between 1.9--2.1 Myr. Indeed, some of the discs were truncated down to 30 $\au$, as they had close flybys of 42 Ori. On the other hand, some discs were only slightly truncated by $\sim$few$\au$ as their flyby event was at a much larger distance, and so the effect of the intense radiation was much reduced.

\begin{figure*}
\centering
\includegraphics[scale=0.55]{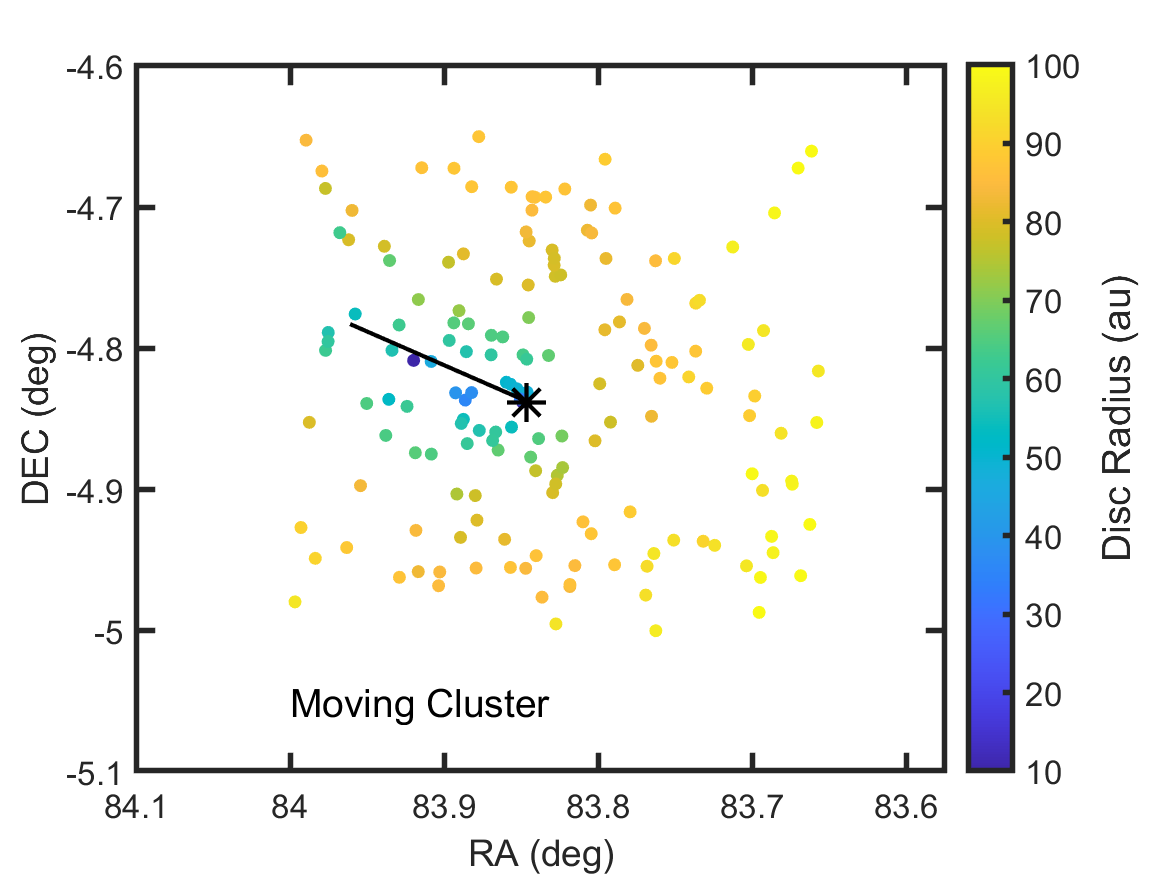}
\hspace{1cm}
\includegraphics[scale=0.55]{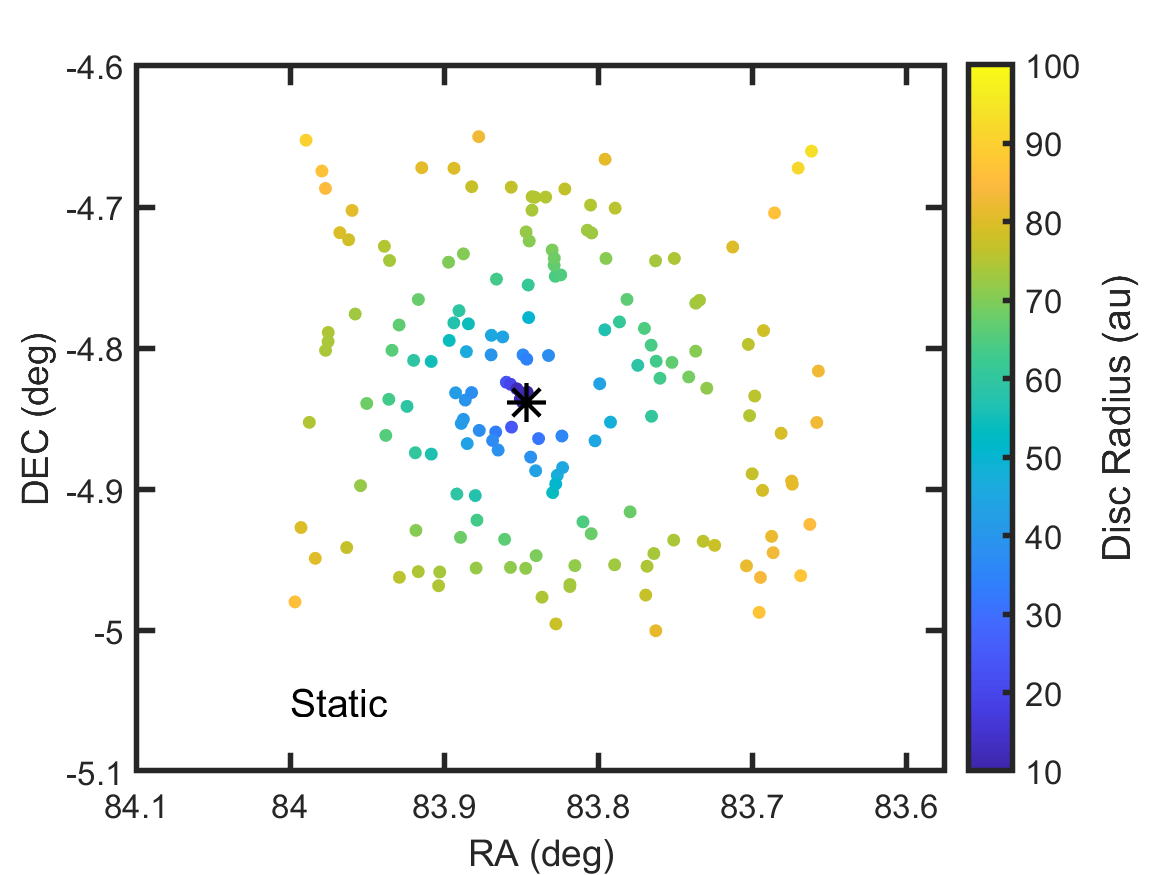}
\caption{Spatial coordinates of the stars in NGC 1977, with the colours showing the simulated disc radii after 2 Myr when all of the stars are moving in the cluster (left-hand panel), and when they are static (right-hand panel). The black star shows the current location of 42 Ori, with the solid black line in the left-hand panel showing its' trajectory over the last 100 Kyr. A movie of the left-hand panel showing the Moving Cluster can be found in the online supplementary material.}
\label{fig:42_ori_map}
\end{figure*}

The variation in disc radii, as discussed in Sect. \ref{sec:observables}, can also be seen in Fig. \ref{fig:42_ori_radii} where before the encounter with 42 Ori, the variance in disc radii was around 25$\au$, whilst after the flybys, the variance increased to $\sim 70\au$. This would imply that as 42 Ori moves through the cluster, then there should be gradients in disc radii in front of and behind the star. Indeed this is seen when overlaying the disc radii on a 2D map of the projected locations of the stars in NGC 1977. This is shown in the left-hand panel of Fig. \ref{fig:42_ori_map} where the colours show the disc radii of the stars as a function of their right ascension and Declination, with the black star signifying 42 Ori, and the black line denoting its' past trajectory. Indeed as 42 Ori moves from the left side of the figure over 100 Kyr, it truncated the discs it comes across, leaving a significant imprint on the disc radii. This valley as shown by the disc radii decreasing with increasing right ascension and decreasing declination along the line of travel represents a 30--40$\au$ difference, which may be seen in future observations. For example 20\,au is equivalent to 0.05 arcseconds at the 400\,pc distance of NGC 1977, which would be resolvable by facilities such as ALMA, JWST NIRCam short wavelength filters and VLT/MUSE Narrow Field Mode. Additionally, when comparing the disc radii in front of 42 Ori, to that behind it, the difference is even greater, in line with the predictions made in Sect. \ref{sec:observables}. These trends are starkly different to those seen in the right-hand panel of Fig. \ref{fig:42_ori_map}, where disc radii are again overlaid on a 2D map, but here the stars are assumed to be static in space, i.e. they would be comoving. For the discs now, they experience the same strong UV fields for their entire lifetimes, typically resulting in smaller discs after 2Myr. Indeed, the disc radii are smallest for the stars close to 42 Ori, before then increasing omnidirectionally away from 42 Ori, a significantly different signature to if 42 Ori is moving through the cluster. Future observations of NGC 1977 should therefore be able to differentiate between the scenarios, and give insights into the dynamical history of the region.

\section{Discussion and Conclusions}
\label{sec:conc}

In this work we have examined the effects that a flyby of a massive star has on the properties of protoplanetary discs and their subsequent evolution. These flybys occur as massive stars move through a stellar cluster, varying the levels of radiation they impart on protoplanetary discs. We modelled this through a hyperbolic tangent function that smoothly increased and then decreased the UV radiation field from a base level to a maximum value, representative of the maximal effects of the flybys. Wit this varying UV field, over time-scales comparable with observed flyby time-scales, we then explored the effects of the flybys on protoplanetary disc radii. We summarise our main results below:

(1) The increase in radiation from the flyby stars acts to rapidly truncate the disc across the flyby time-scale. After the flyby has occurred, the discs then viscously rebound before gradually decreasing in size again as they evolve under their normal conditions. However, even with this rebound, they are always much smaller and less massive than discs that did not experience a flyby, and as such they may appear to look older than they are when including them in disc evolution studies.

(2) Whilst the specific timing of a flyby event is shown not to have a large impact, the maximum UV field strength and the time-scale of the flyby are shown to have significant effects on how much truncation discs undergo. With longer flyby time-scales and/or larger maximum UV field strengths, more radiation is imparted on to the discs, driving larger photoevaporative winds, that ultimately deplete the discs at a faster rate. For longer flybys of the most massive stars, this can result in disc lifetimes reduced by millions of years, and disc radii drastically reduced compared to discs undergoing weaker flyby events, or no flybys at all.

(3) As a massive star moves through a cluster, its' flyby of different stars in that cluster will occur at different times of their lifetimes. Naturally this results in stars in front of the flyby star not having their event yet, whilst those behind it, have already undergone a flyby. This difference in flyby times within a cluster therefore results in different disc radii of the stars along the trajectory of the flyby star, with more massive discs in front of the flyby star, and smaller ones behind. Depending on the maximum UV strength, flyby time-scale, and the differential ages of the stars in the cluster, the differences in radii around a flyby star are found to vary from between 10--50$\au$.

(4) As well as variances in stellar ages affecting the the differences in radii around a flyby star, we also find that the stellar mass and initial compactness of the disc, denoted by $r_{\rm c}$, also induce significant variance. When changing the stellar mass between 0.3--1$\msun$ the differences in disc radii around a flyby star can vary by between 10--40 $\au$, whilst changes in the initial disc compactness also allows for differences in disc radii between 20--40 $\au$. These are comparable to the differences found for stars of different ages in front of and behind a flyby star, highlighting the significant degeneracy that exists when using the differences in disc radii around a moving massive star to explore the properties of protoplanetary discs and their evolution.

(5) Finally, we apply our models to the NGC 1977 cluster where the massive star, 42 Ori, is predicted to be passing through, and therefore having flyby events with the stars in the cluster. Our results show that the stars in the cluster are significantly impacted by 42 Ori passing through, with disc radii differing by 40$\au$ in size, depending on where the stars are in relation to 42 Ori. Indeed, those stars behind 42 Ori, along its' past trajectory, show the most extreme differences, where a valley of irradiation is found. Additionally those stars in front of 42 Ori, remain relatively untouched for the time being, but will undergo their flyby events in the near future. With them being currently unaffected, this gives an additional gradient in disc radii from in front of 42 Ori to behind it. Future observations of disc radii around the stars in the vicinity of 42 Ori should be able to observe such features and confirm these theoretical expectations.

In summary, we have shown that flyby events can significantly reduce the mass and size of protoplanetary discs, leaving observable signatures that future missions and observing campaigns should find. The main observable feature will be in the disc radii around a massive star. If the star was static compared to its residing cluster, then it would be expected that the changes in disc radii would be uniform in respect to the azimuthal direction around the massive star. However, if the massive star was moving through the cluster, then a valley would appear behind the massive star, along it's direction of travel, showcasing the effects of the flybys. Observations of this valley, and quantifying the differences in disc radii for those discs, will be useful in understanding disc evolution processes. Only then by understanding and accurately modelling the effects of the environment on evolving protoplanetary discs, will we be able to begin effectively exploring how protoplanetary discs evolve, and subsequently how planets and planetary systems form in stellar clusters.

\section*{Data Availability}
The data underlying this article will be shared on reasonable request to the corresponding author.

\section*{Acknowledgements}
The authors thank the anonymous referee for providing useful and interesting comments that improved the paper.
GALC acknowledges funding from the UKRI/STFC grant ST/X000931/1.
TJH acknowledges UKRI guaranteed funding for a Horizon Europe ERC consolidator grant (EP/Y024710/1) and a Royal Society Dorothy Hodgkin Fellowship. 
This research utilised Queen Mary's Apocrita HPC facility, supported by QMUL Research-IT (http://doi.org/10.5281/zenodo.438045).
JSK acknowledges NASA’s Nexus for Exoplanet System Science (NExSS) research coordination network sponsored by NASA’s Science Mission Directorate and project “Alien Earths” funded under Agreement No. 80NSSC21K0593. PAH and TCK were supported in part through the Arizona NASA Space Grant Consortium, Cooperative Agreement 80NSSC20M0041.

\bibliographystyle{mnras}
\bibliography{references}{}

\begin{thebibliography}{}
\makeatletter
\relax
\def\mn@urlcharsother{\let\do\@makeother \do\$\do\&\do\#\do\^\do\_\do\%\do\~}
\def\mn@doi{\begingroup\mn@urlcharsother \@ifnextchar [ {\mn@doi@} {\mn@doi@[]}}
\def\mn@doi@[#1]#2{\def\@tempa{#1}\ifx\@tempa\@empty \href {http://dx.doi.org/#2} {doi:#2}\else \href {http://dx.doi.org/#2} {#1}\fi \endgroup}
\def\mn@eprint#1#2{\mn@eprint@#1:#2::\@nil}
\def\mn@eprint@arXiv#1{\href {http://arxiv.org/abs/#1} {{\tt arXiv:#1}}}
\def\mn@eprint@dblp#1{\href {http://dblp.uni-trier.de/rec/bibtex/#1.xml} {dblp:#1}}
\def\mn@eprint@#1:#2:#3:#4\@nil{\def\@tempa {#1}\def\@tempb {#2}\def\@tempc {#3}\ifx \@tempc \@empty \let \@tempc \@tempb \let \@tempb \@tempa \fi \ifx \@tempb \@empty \def\@tempb {arXiv}\fi \@ifundefined {mn@eprint@\@tempb}{\@tempb:\@tempc}{\expandafter \expandafter \csname mn@eprint@\@tempb\endcsname \expandafter{\@tempc}}}

\bibitem[\protect\citeauthoryear{{Adams}, {Hollenbach}, {Laughlin}  \& {Gorti}}{{Adams} et~al.}{2004}]{Adams2004}
{Adams} F.~C.,  {Hollenbach} D.,  {Laughlin} G.,   {Gorti} U.,  2004, \mn@doi [\apj] {10.1086/421989}, \href {https://ui.adsabs.harvard.edu/abs/2004ApJ...611..360A} {611, 360}

\bibitem[\protect\citeauthoryear{{Anania}, {Winter}, {Rosotti}, {Vioque}, {Zari}, {Pantaleoni Gonz{\'a}lez}  \& {Testi}}{{Anania} et~al.}{2025}]{Anania25}
{Anania} R.,  {Winter} A.~J.,  {Rosotti} G.,  {Vioque} M.,  {Zari} E.,  {Pantaleoni Gonz{\'a}lez} M.,   {Testi} L.,  2025, \mn@doi [\aap] {10.1051/0004-6361/202453011}, \href {https://ui.adsabs.harvard.edu/abs/2025A&A...695A..74A} {695, A74}

\bibitem[\protect\citeauthoryear{{Ansdell}, {Williams}, {Manara}, {Miotello}, {Facchini}, {van der Marel}, {Testi}  \& {van Dishoeck}}{{Ansdell} et~al.}{2017}]{Ansdell17}
{Ansdell} M.,  {Williams} J.~P.,  {Manara} C.~F.,  {Miotello} A.,  {Facchini} S.,  {van der Marel} N.,  {Testi} L.,   {van Dishoeck} E.~F.,  2017, \mn@doi [\aj] {10.3847/1538-3881/aa69c0}, \href {https://ui.adsabs.harvard.edu/abs/2017AJ....153..240A} {153, 240}

\bibitem[\protect\citeauthoryear{{Ansdell} et~al.,}{{Ansdell} et~al.}{2018}]{Ansdell18}
{Ansdell} M.,  et~al., 2018, \mn@doi [\apj] {10.3847/1538-4357/aab890}, \href {https://ui.adsabs.harvard.edu/abs/2018ApJ...859...21A} {859, 21}

\bibitem[\protect\citeauthoryear{{Aru} et~al.,}{{Aru} et~al.}{2024}]{Aru24}
{Aru} M.~L.,  et~al., 2024, \mn@doi [\aap] {10.1051/0004-6361/202349004}, \href {https://ui.adsabs.harvard.edu/abs/2024A&A...687A..93A} {687, A93}

\bibitem[\protect\citeauthoryear{{Ballabio}, {Haworth}  \& {Henney}}{{Ballabio} et~al.}{2023}]{Ballabio23}
{Ballabio} G.,  {Haworth} T.~J.,   {Henney} W.~J.,  2023, \mn@doi [\mnras] {10.1093/mnras/stac3467}, \href {https://ui.adsabs.harvard.edu/abs/2023MNRAS.518.5563B} {518, 5563}

\bibitem[\protect\citeauthoryear{{Ballering} et~al.,}{{Ballering} et~al.}{2023}]{Ballering23}
{Ballering} N.~P.,  et~al., 2023, \mn@doi [\apj] {10.3847/1538-4357/ace901}, \href {https://ui.adsabs.harvard.edu/abs/2023ApJ...954..127B} {954, 127}

\bibitem[\protect\citeauthoryear{{Bern{\'e}} et~al.,}{{Bern{\'e}} et~al.}{2024}]{2024Sci...383..988B}
{Bern{\'e}} O.,  et~al., 2024, \mn@doi [Science] {10.1126/science.adh2861}, \href {https://ui.adsabs.harvard.edu/abs/2024Sci...383..988B} {383, 988}

\bibitem[\protect\citeauthoryear{{Carretero-Castrillo}, {Benaglia}, {Paredes}  \& {Rib{\'o}}}{{Carretero-Castrillo} et~al.}{2025}]{2025arXiv250202658C}
{Carretero-Castrillo} M.,  {Benaglia} P.,  {Paredes} J.~M.,   {Rib{\'o}} M.,  2025, arXiv e-prints, \href {https://ui.adsabs.harvard.edu/abs/2025arXiv250202658C} {p. arXiv:2502.02658}

\bibitem[\protect\citeauthoryear{{Clarke}}{{Clarke}}{2007}]{Clarke2007}
{Clarke} C.~J.,  2007, \mn@doi [\mnras] {10.1111/j.1365-2966.2007.11547.x}, \href {https://ui.adsabs.harvard.edu/abs/2007MNRAS.376.1350C} {376, 1350}

\bibitem[\protect\citeauthoryear{{Clarke} \& {Pringle}}{{Clarke} \& {Pringle}}{1993}]{1993MNRAS.261..190C}
{Clarke} C.~J.,  {Pringle} J.~E.,  1993, \mn@doi [\mnras] {10.1093/mnras/261.1.190}, \href {https://ui.adsabs.harvard.edu/abs/1993MNRAS.261..190C} {261, 190}

\bibitem[\protect\citeauthoryear{{Clarke}, {Gendrin}  \& {Sotomayor}}{{Clarke} et~al.}{2001}]{Clarke2001}
{Clarke} C.~J.,  {Gendrin} A.,   {Sotomayor} M.,  2001, \mn@doi [\mnras] {10.1046/j.1365-8711.2001.04891.x}, \href {http://adsabs.harvard.edu/abs/2001MNRAS.328..485C} {328, 485}

\bibitem[\protect\citeauthoryear{{Coleman}}{{Coleman}}{2021}]{Coleman21}
{Coleman} G. A.~L.,  2021, \mn@doi [\mnras] {10.1093/mnras/stab1904}, \href {https://ui.adsabs.harvard.edu/abs/2021MNRAS.506.3596C} {506, 3596}

\bibitem[\protect\citeauthoryear{{Coleman} \& {Haworth}}{{Coleman} \& {Haworth}}{2020}]{ColemanHaworth20}
{Coleman} G. A.~L.,  {Haworth} T.~J.,  2020, \mn@doi [\mnras] {10.1093/mnrasl/slaa098}, \href {https://ui.adsabs.harvard.edu/abs/2020MNRAS.496L.111C} {496, L111}

\bibitem[\protect\citeauthoryear{{Coleman} \& {Haworth}}{{Coleman} \& {Haworth}}{2022}]{Coleman22}
{Coleman} G. A.~L.,  {Haworth} T.~J.,  2022, \mn@doi [MNRAS] {10.1093/mnras/stac1513}, \href {https://ui.adsabs.harvard.edu/abs/2022MNRAS.514.2315C} {514, 2315}

\bibitem[\protect\citeauthoryear{{Coleman}, {Mroueh}  \& {Haworth}}{{Coleman} et~al.}{2024}]{Coleman24MHD}
{Coleman} G. A.~L.,  {Mroueh} J.~K.,   {Haworth} T.~J.,  2024, \mn@doi [\mnras] {10.1093/mnras/stad3692}, \href {https://ui.adsabs.harvard.edu/abs/2024MNRAS.527.7588C} {527, 7588}

\bibitem[\protect\citeauthoryear{{Coleman}, {Haworth}  \& {Qiao}}{{Coleman} et~al.}{2025}]{Coleman25}
{Coleman} G. A.~L.,  {Haworth} T.~J.,   {Qiao} L.,  2025, \mn@doi [\mnras] {10.1093/mnras/staf555}, \href {https://ui.adsabs.harvard.edu/abs/2025MNRAS.539.1190C} {539, 1190}

\bibitem[\protect\citeauthoryear{{Concha-Ram{\'\i}rez}, {Wilhelm}, {Portegies Zwart}  \& {Haworth}}{{Concha-Ram{\'\i}rez} et~al.}{2019}]{ConchaRamirez19}
{Concha-Ram{\'\i}rez} F.,  {Wilhelm} M. J.~C.,  {Portegies Zwart} S.,   {Haworth} T.~J.,  2019, \mn@doi [\mnras] {10.1093/mnras/stz2973}, \href {https://ui.adsabs.harvard.edu/abs/2019MNRAS.490.5678C} {490, 5678}

\bibitem[\protect\citeauthoryear{{Concha-Ram{\'\i}rez}, {Wilhelm}, {Portegies Zwart}, {van Terwisga}  \& {Hacar}}{{Concha-Ram{\'\i}rez} et~al.}{2021}]{ConchaRamirez21}
{Concha-Ram{\'\i}rez} F.,  {Wilhelm} M. J.~C.,  {Portegies Zwart} S.,  {van Terwisga} S.~E.,   {Hacar} A.,  2021, \mn@doi [\mnras] {10.1093/mnras/staa3669}, \href {https://ui.adsabs.harvard.edu/abs/2021MNRAS.501.1782C} {501, 1782}

\bibitem[\protect\citeauthoryear{{Cuello}, {M{\'e}nard}  \& {Price}}{{Cuello} et~al.}{2023}]{2023EPJP..138...11C}
{Cuello} N.,  {M{\'e}nard} F.,   {Price} D.~J.,  2023, \mn@doi [European Physical Journal Plus] {10.1140/epjp/s13360-022-03602-w}, \href {https://ui.adsabs.harvard.edu/abs/2023EPJP..138...11C} {138, 11}

\bibitem[\protect\citeauthoryear{{D'Angelo} \& {Marzari}}{{D'Angelo} \& {Marzari}}{2012}]{Dangelo12}
{D'Angelo} G.,  {Marzari} F.,  2012, \mn@doi [\apj] {10.1088/0004-637X/757/1/50}, \href {http://adsabs.harvard.edu/abs/2012ApJ...757...50D} {757, 50}

\bibitem[\protect\citeauthoryear{{Eatson}, {Parker}  \& {Lichtenberg}}{{Eatson} et~al.}{2024}]{2024arXiv241017163E}
{Eatson} J.~W.,  {Parker} R.~J.,   {Lichtenberg} T.,  2024, \mn@doi [arXiv e-prints] {10.48550/arXiv.2410.17163}, \href {https://ui.adsabs.harvard.edu/abs/2024arXiv241017163E} {p. arXiv:2410.17163}

\bibitem[\protect\citeauthoryear{{Eisner} et~al.,}{{Eisner} et~al.}{2018}]{Eisner18}
{Eisner} J.~A.,  et~al., 2018, \mn@doi [\apj] {10.3847/1538-4357/aac3e2}, \href {https://ui.adsabs.harvard.edu/abs/2018ApJ...860...77E} {860, 77}

\bibitem[\protect\citeauthoryear{{Ercolano}, {Picogna}, {Monsch}, {Drake}  \& {Preibisch}}{{Ercolano} et~al.}{2021}]{Ercolano21}
{Ercolano} B.,  {Picogna} G.,  {Monsch} K.,  {Drake} J.~J.,   {Preibisch} T.,  2021, \mn@doi [\mnras] {10.1093/mnras/stab2590}, \href {https://ui.adsabs.harvard.edu/abs/2021MNRAS.508.1675E} {508, 1675}

\bibitem[\protect\citeauthoryear{{Fujii} \& {Portegies Zwart}}{{Fujii} \& {Portegies Zwart}}{2011}]{2011Sci...334.1380F}
{Fujii} M.~S.,  {Portegies Zwart} S.,  2011, \mn@doi [Science] {10.1126/science.1211927}, \href {https://ui.adsabs.harvard.edu/abs/2011Sci...334.1380F} {334, 1380}

\bibitem[\protect\citeauthoryear{{G{\'a}rate}, {Pinilla}, {Haworth}  \& {Facchini}}{{G{\'a}rate} et~al.}{2024}]{Garate2024}
{G{\'a}rate} M.,  {Pinilla} P.,  {Haworth} T.~J.,   {Facchini} S.,  2024, \mn@doi [\aap] {10.1051/0004-6361/202347850}, \href {https://ui.adsabs.harvard.edu/abs/2024A&A...681A..84G} {681, A84}

\bibitem[\protect\citeauthoryear{{Gorti}, {Dullemond}  \& {Hollenbach}}{{Gorti} et~al.}{2009}]{Gorti09}
{Gorti} U.,  {Dullemond} C.~P.,   {Hollenbach} D.,  2009, \mn@doi [\apj] {10.1088/0004-637X/705/2/1237}, \href {https://ui.adsabs.harvard.edu/abs/2009ApJ...705.1237G} {705, 1237}

\bibitem[\protect\citeauthoryear{{Gorti}, {Hollenbach}  \& {Dullemond}}{{Gorti} et~al.}{2015}]{Gorti15}
{Gorti} U.,  {Hollenbach} D.,   {Dullemond} C.~P.,  2015, \mn@doi [\apj] {10.1088/0004-637X/804/1/29}, \href {https://ui.adsabs.harvard.edu/abs/2015ApJ...804...29G} {804, 29}

\bibitem[\protect\citeauthoryear{{Guarcello} et~al.,}{{Guarcello} et~al.}{2016}]{Guarcello16}
{Guarcello} M.~G.,  et~al., 2016, \mn@doi [arXiv e-prints] {10.48550/arXiv.1605.01773}, \href {https://ui.adsabs.harvard.edu/abs/2016arXiv160501773G} {p. arXiv:1605.01773}

\bibitem[\protect\citeauthoryear{{Gupta}, {Miotello}, {Williams}, {Birnstiel}, {Kuffmeier}  \& {Yen}}{{Gupta} et~al.}{2024}]{2024A&A...683A.133G}
{Gupta} A.,  {Miotello} A.,  {Williams} J.~P.,  {Birnstiel} T.,  {Kuffmeier} M.,   {Yen} H.-W.,  2024, \mn@doi [\aap] {10.1051/0004-6361/202348007}, \href {https://ui.adsabs.harvard.edu/abs/2024A&A...683A.133G} {683, A133}

\bibitem[\protect\citeauthoryear{{Hallatt} \& {Lee}}{{Hallatt} \& {Lee}}{2025}]{Hallatt2025}
{Hallatt} T.,  {Lee} E.~J.,  2025, \mn@doi [\apj] {10.3847/1538-4357/ad9aa1}, \href {https://ui.adsabs.harvard.edu/abs/2025ApJ...979..120H} {979, 120}

\bibitem[\protect\citeauthoryear{{Haworth} \& {Clarke}}{{Haworth} \& {Clarke}}{2019}]{Haworth19}
{Haworth} T.~J.,  {Clarke} C.~J.,  2019, \mn@doi [\mnras] {10.1093/mnras/stz706}, \href {https://ui.adsabs.harvard.edu/abs/2019MNRAS.485.3895H} {485, 3895}

\bibitem[\protect\citeauthoryear{{Haworth}, {Facchini}, {Clarke}  \& {Cleeves}}{{Haworth} et~al.}{2017}]{Haworth17}
{Haworth} T.~J.,  {Facchini} S.,  {Clarke} C.~J.,   {Cleeves} L.~I.,  2017, \mn@doi [\mnras] {10.1093/mnrasl/slx037}, \href {https://ui.adsabs.harvard.edu/abs/2017MNRAS.468L.108H} {468, L108}

\bibitem[\protect\citeauthoryear{{Haworth}, {Facchini}, {Clarke}  \& {Mohanty}}{{Haworth} et~al.}{2018a}]{2018MNRAS.475.5460H}
{Haworth} T.~J.,  {Facchini} S.,  {Clarke} C.~J.,   {Mohanty} S.,  2018a, \mn@doi [\mnras] {10.1093/mnras/sty168}, \href {https://ui.adsabs.harvard.edu/abs/2018MNRAS.475.5460H} {475, 5460}

\bibitem[\protect\citeauthoryear{{Haworth}, {Clarke}, {Rahman}, {Winter}  \& {Facchini}}{{Haworth} et~al.}{2018b}]{Haworth18}
{Haworth} T.~J.,  {Clarke} C.~J.,  {Rahman} W.,  {Winter} A.~J.,   {Facchini} S.,  2018b, \mn@doi [\mnras] {10.1093/mnras/sty2323}, \href {https://ui.adsabs.harvard.edu/abs/2018MNRAS.481..452H} {481, 452}

\bibitem[\protect\citeauthoryear{{Haworth}, {Kim}, {Winter}, {Hines}, {Clarke}, {Sellek}, {Ballabio}  \& {Stapelfeldt}}{{Haworth} et~al.}{2021}]{Haworth21}
{Haworth} T.~J.,  {Kim} J.~S.,  {Winter} A.~J.,  {Hines} D.~C.,  {Clarke} C.~J.,  {Sellek} A.~D.,  {Ballabio} G.,   {Stapelfeldt} K.~R.,  2021, \mn@doi [\mnras] {10.1093/mnras/staa3918}, \href {https://ui.adsabs.harvard.edu/abs/2021MNRAS.501.3502H} {501, 3502}

\bibitem[\protect\citeauthoryear{{Haworth}, {Coleman}, {Qiao}, {Sellek}  \& {Askari}}{{Haworth} et~al.}{2023}]{Haworth23}
{Haworth} T.~J.,  {Coleman} G. A.~L.,  {Qiao} L.,  {Sellek} A.~D.,   {Askari} K.,  2023, \mn@doi [\mnras] {10.1093/mnras/stad3054}, \href {https://ui.adsabs.harvard.edu/abs/2023MNRAS.526.4315H} {526, 4315}

\bibitem[\protect\citeauthoryear{{Henney} \& {O'Dell}}{{Henney} \& {O'Dell}}{1999}]{Henney1999}
{Henney} W.~J.,  {O'Dell} C.~R.,  1999, \mn@doi [\aj] {10.1086/301087}, \href {https://ui.adsabs.harvard.edu/abs/1999AJ....118.2350H} {118, 2350}

\bibitem[\protect\citeauthoryear{{Holden}, {Landis}, {Spitzig}  \& {Adams}}{{Holden} et~al.}{2011}]{2011PASP..123...14H}
{Holden} L.,  {Landis} E.,  {Spitzig} J.,   {Adams} F.~C.,  2011, \mn@doi [\pasp] {10.1086/658081}, \href {https://ui.adsabs.harvard.edu/abs/2011PASP..123...14H} {123, 14}

\bibitem[\protect\citeauthoryear{{Huang}, {Portegies Zwart}  \& {Wilhelm}}{{Huang} et~al.}{2024}]{Huang24}
{Huang} S.,  {Portegies Zwart} S.,   {Wilhelm} M. J.~C.,  2024, \mn@doi [\aap] {10.1051/0004-6361/202451051}, \href {https://ui.adsabs.harvard.edu/abs/2024A&A...689A.338H} {689, A338}

\bibitem[\protect\citeauthoryear{{Johnstone}, {Hollenbach}  \& {Bally}}{{Johnstone} et~al.}{1998}]{Johnstone98}
{Johnstone} D.,  {Hollenbach} D.,   {Bally} J.,  1998, \mn@doi [\apj] {10.1086/305658}, \href {https://ui.adsabs.harvard.edu/abs/1998ApJ...499..758J} {499, 758}

\bibitem[\protect\citeauthoryear{{Kim}, {Clarke}, {Fang}  \& {Facchini}}{{Kim} et~al.}{2016}]{Kim2016}
{Kim} J.~S.,  {Clarke} C.~J.,  {Fang} M.,   {Facchini} S.,  2016, \mn@doi [\apjl] {10.3847/2041-8205/826/1/L15}, \href {https://ui.adsabs.harvard.edu/abs/2016ApJ...826L..15K} {826, L15}

\bibitem[\protect\citeauthoryear{{Komaki}, {Fukuhara}, {Suzuki}  \& {Yoshida}}{{Komaki} et~al.}{2023}]{Komaki23}
{Komaki} A.,  {Fukuhara} S.,  {Suzuki} T.~K.,   {Yoshida} N.,  2023, \mn@doi [arXiv e-prints] {10.48550/arXiv.2304.13316}, \href {https://ui.adsabs.harvard.edu/abs/2023arXiv230413316K} {p. arXiv:2304.13316}

\bibitem[\protect\citeauthoryear{{Lichtenberg}, {Golabek}, {Gerya}  \& {Meyer}}{{Lichtenberg} et~al.}{2016}]{2016Icar..274..350L}
{Lichtenberg} T.,  {Golabek} G.~J.,  {Gerya} T.~V.,   {Meyer} M.~R.,  2016, \mn@doi [\icarus] {10.1016/j.icarus.2016.03.004}, \href {https://ui.adsabs.harvard.edu/abs/2016Icar..274..350L} {274, 350}

\bibitem[\protect\citeauthoryear{{Lynden-Bell} \& {Pringle}}{{Lynden-Bell} \& {Pringle}}{1974}]{Lynden-BellPringle1974}
{Lynden-Bell} D.,  {Pringle} J.~E.,  1974, \mnras, \href {http://adsabs.harvard.edu/abs/1974MNRAS.168..603L} {168, 603}

\bibitem[\protect\citeauthoryear{{Mann} et~al.,}{{Mann} et~al.}{2014}]{Mann14}
{Mann} R.~K.,  et~al., 2014, \mn@doi [\apj] {10.1088/0004-637X/784/1/82}, \href {https://ui.adsabs.harvard.edu/abs/2014ApJ...784...82M} {784, 82}

\bibitem[\protect\citeauthoryear{{Matsuyama}, {Johnstone}  \& {Hartmann}}{{Matsuyama} et~al.}{2003}]{Matsuyama03}
{Matsuyama} I.,  {Johnstone} D.,   {Hartmann} L.,  2003, \mn@doi [\apj] {10.1086/344638}, \href {https://ui.adsabs.harvard.edu/abs/2003ApJ...582..893M} {582, 893}

\bibitem[\protect\citeauthoryear{{Mauc{\'o}} et~al.,}{{Mauc{\'o}} et~al.}{2023}]{2023A&A...680C...1M}
{Mauc{\'o}} K.,  et~al., 2023, \mn@doi [\aap] {10.1051/0004-6361/202348748e}, \href {https://ui.adsabs.harvard.edu/abs/2023A&A...680C...1M} {680, C1}

\bibitem[\protect\citeauthoryear{{Mesa-Delgado}, {N{\'u}{\~n}ez-D{\'\i}az}, {Esteban}, {Garc{\'\i}a-Rojas}, {Flores-Fajardo}, {L{\'o}pez-Mart{\'\i}n}, {Tsamis}  \& {Henney}}{{Mesa-Delgado} et~al.}{2012}]{MesaDelgado12}
{Mesa-Delgado} A.,  {N{\'u}{\~n}ez-D{\'\i}az} M.,  {Esteban} C.,  {Garc{\'\i}a-Rojas} J.,  {Flores-Fajardo} N.,  {L{\'o}pez-Mart{\'\i}n} L.,  {Tsamis} Y.~G.,   {Henney} W.~J.,  2012, \mn@doi [\mnras] {10.1111/j.1365-2966.2012.21230.x}, \href {https://ui.adsabs.harvard.edu/abs/2012MNRAS.426..614M} {426, 614}

\bibitem[\protect\citeauthoryear{{Nakatani}, {Hosokawa}, {Yoshida}, {Nomura}  \& {Kuiper}}{{Nakatani} et~al.}{2018}]{Nakatani18}
{Nakatani} R.,  {Hosokawa} T.,  {Yoshida} N.,  {Nomura} H.,   {Kuiper} R.,  2018, \mn@doi [\apj] {10.3847/1538-4357/aad9fd}, \href {https://ui.adsabs.harvard.edu/abs/2018ApJ...865...75N} {865, 75}

\bibitem[\protect\citeauthoryear{{Owen}, {Ercolano}, {Clarke}  \& {Alexander}}{{Owen} et~al.}{2010}]{Owen10}
{Owen} J.~E.,  {Ercolano} B.,  {Clarke} C.~J.,   {Alexander} R.~D.,  2010, \mn@doi [\mnras] {10.1111/j.1365-2966.2009.15771.x}, \href {https://ui.adsabs.harvard.edu/abs/2010MNRAS.401.1415O} {401, 1415}

\bibitem[\protect\citeauthoryear{{Owen}, {Clarke}  \& {Ercolano}}{{Owen} et~al.}{2012}]{Owen12}
{Owen} J.~E.,  {Clarke} C.~J.,   {Ercolano} B.,  2012, \mn@doi [\mnras] {10.1111/j.1365-2966.2011.20337.x}, \href {https://ui.adsabs.harvard.edu/abs/2012MNRAS.422.1880O} {422, 1880}

\bibitem[\protect\citeauthoryear{{Parker}, {Alcock}, {Nicholson}, {Pani{\'c}}  \& {Goodwin}}{{Parker} et~al.}{2021}]{2021ApJ...913...95P}
{Parker} R.~J.,  {Alcock} H.~L.,  {Nicholson} R.~B.,  {Pani{\'c}} O.,   {Goodwin} S.~P.,  2021, \mn@doi [\apj] {10.3847/1538-4357/abf4cc}, \href {https://ui.adsabs.harvard.edu/abs/2021ApJ...913...95P} {913, 95}

\bibitem[\protect\citeauthoryear{{Pascucci}, {Cabrit}, {Edwards}, {Gorti}, {Gressel}  \& {Suzuki}}{{Pascucci} et~al.}{2023}]{2023ASPC..534..567P}
{Pascucci} I.,  {Cabrit} S.,  {Edwards} S.,  {Gorti} U.,  {Gressel} O.,   {Suzuki} T.~K.,  2023, in {Inutsuka} S.,  {Aikawa} Y.,  {Muto} T.,  {Tomida} K.,   {Tamura} M.,  eds,  Astronomical Society of the Pacific Conference Series Vol. 534, Protostars and Planets VII. p.~567 (\mn@eprint {arXiv} {2203.10068}), \mn@doi{10.48550/arXiv.2203.10068}

\bibitem[\protect\citeauthoryear{{Picogna}, {Ercolano}, {Owen}  \& {Weber}}{{Picogna} et~al.}{2019}]{Picogna19}
{Picogna} G.,  {Ercolano} B.,  {Owen} J.~E.,   {Weber} M.~L.,  2019, \mn@doi [\mnras] {10.1093/mnras/stz1166}, \href {https://ui.adsabs.harvard.edu/abs/2019MNRAS.487..691P} {487, 691}

\bibitem[\protect\citeauthoryear{{Picogna}, {Ercolano}  \& {Espaillat}}{{Picogna} et~al.}{2021}]{Picogna21}
{Picogna} G.,  {Ercolano} B.,   {Espaillat} C.~C.,  2021, \mn@doi [\mnras] {10.1093/mnras/stab2883}, \href {https://ui.adsabs.harvard.edu/abs/2021MNRAS.508.3611P} {508, 3611}

\bibitem[\protect\citeauthoryear{{Pineda}, {Segura-Cox}, {Caselli}, {Cunningham}, {Zhao}, {Schmiedeke}, {Maureira}  \& {Neri}}{{Pineda} et~al.}{2020}]{2020NatAs...4.1158P}
{Pineda} J.~E.,  {Segura-Cox} D.,  {Caselli} P.,  {Cunningham} N.,  {Zhao} B.,  {Schmiedeke} A.,  {Maureira} M.~J.,   {Neri} R.,  2020, \mn@doi [Nature Astronomy] {10.1038/s41550-020-1150-z}, \href {https://ui.adsabs.harvard.edu/abs/2020NatAs...4.1158P} {4, 1158}

\bibitem[\protect\citeauthoryear{{Planet formation environments collaboration} et~al.,}{{Planet formation environments collaboration} et~al.}{2025}]{PlanetFormationCollaboration}
{Planet formation environments collaboration} et~al., 2025, \mn@doi [arXiv e-prints] {10.48550/arXiv.2502.12255}, \href {https://ui.adsabs.harvard.edu/abs/2025arXiv250212255P} {p. arXiv:2502.12255}

\bibitem[\protect\citeauthoryear{{Qiao}, {Haworth}, {Sellek}  \& {Ali}}{{Qiao} et~al.}{2022}]{Qiao22}
{Qiao} L.,  {Haworth} T.~J.,  {Sellek} A.~D.,   {Ali} A.~A.,  2022, \mn@doi [\mnras] {10.1093/mnras/stac684}, \href {https://ui.adsabs.harvard.edu/abs/2022MNRAS.512.3788Q} {512, 3788}

\bibitem[\protect\citeauthoryear{{Qiao}, {Coleman}  \& {Haworth}}{{Qiao} et~al.}{2023}]{Qiao23}
{Qiao} L.,  {Coleman} G. A.~L.,   {Haworth} T.~J.,  2023, \mn@doi [\mnras] {10.1093/mnras/stad944}, \href {https://ui.adsabs.harvard.edu/abs/2023MNRAS.522.1939Q} {522, 1939}

\bibitem[\protect\citeauthoryear{{Reiter} \& {Parker}}{{Reiter} \& {Parker}}{2022}]{ParkerReiter22}
{Reiter} M.,  {Parker} R.~J.,  2022, \mn@doi [European Physical Journal Plus] {10.1140/epjp/s13360-022-03265-7}, \href {https://ui.adsabs.harvard.edu/abs/2022EPJP..137.1071R} {137, 1071}

\bibitem[\protect\citeauthoryear{{Renzo} et~al.,}{{Renzo} et~al.}{2019}]{2019A&A...624A..66R}
{Renzo} M.,  et~al., 2019, \mn@doi [\aap] {10.1051/0004-6361/201833297}, \href {https://ui.adsabs.harvard.edu/abs/2019A&A...624A..66R} {624, A66}

\bibitem[\protect\citeauthoryear{{Sellek}, {Booth}  \& {Clarke}}{{Sellek} et~al.}{2020}]{Sellek20}
{Sellek} A.~D.,  {Booth} R.~A.,   {Clarke} C.~J.,  2020, \mn@doi [\mnras] {10.1093/mnras/stz3528}, \href {https://ui.adsabs.harvard.edu/abs/2020MNRAS.492.1279S} {492, 1279}

\bibitem[\protect\citeauthoryear{{Sellek}, {Clarke}  \& {Ercolano}}{{Sellek} et~al.}{2022}]{Sellek22}
{Sellek} A.~D.,  {Clarke} C.~J.,   {Ercolano} B.,  2022, \mn@doi [\mnras] {10.1093/mnras/stac1148}, \href {https://ui.adsabs.harvard.edu/abs/2022MNRAS.514..535S} {514, 535}

\bibitem[\protect\citeauthoryear{{Sellek}, {Grassi}, {Picogna}, {Rab}, {Clarke}  \& {Ercolano}}{{Sellek} et~al.}{2024}]{Sellek24}
{Sellek} A.~D.,  {Grassi} T.,  {Picogna} G.,  {Rab} C.,  {Clarke} C.~J.,   {Ercolano} B.,  2024, \mn@doi [arXiv e-prints] {10.48550/arXiv.2408.00848}, \href {https://ui.adsabs.harvard.edu/abs/2024arXiv240800848S} {p. arXiv:2408.00848}

\bibitem[\protect\citeauthoryear{{Shakura} \& {Sunyaev}}{{Shakura} \& {Sunyaev}}{1973}]{Shak}
{Shakura} N.~I.,  {Sunyaev} R.~A.,  1973, \aap, \href {http://adsabs.harvard.edu/abs/1973A%26A....24..337S} {24, 337}

\bibitem[\protect\citeauthoryear{{St{\"o}rzer} \& {Hollenbach}}{{St{\"o}rzer} \& {Hollenbach}}{1998}]{1998ApJ...502L..71S}
{St{\"o}rzer} H.,  {Hollenbach} D.,  1998, \mn@doi [\apjl] {10.1086/311487}, \href {https://ui.adsabs.harvard.edu/abs/1998ApJ...502L..71S} {502, L71}

\bibitem[\protect\citeauthoryear{{St{\"o}rzer} \& {Hollenbach}}{{St{\"o}rzer} \& {Hollenbach}}{1999}]{1999ApJ...515..669S}
{St{\"o}rzer} H.,  {Hollenbach} D.,  1999, \mn@doi [\apj] {10.1086/307055}, \href {https://ui.adsabs.harvard.edu/abs/1999ApJ...515..669S} {515, 669}

\bibitem[\protect\citeauthoryear{{Tabone}, {Rosotti}, {Cridland}, {Armitage}  \& {Lodato}}{{Tabone} et~al.}{2022}]{Tabone22}
{Tabone} B.,  {Rosotti} G.~P.,  {Cridland} A.~J.,  {Armitage} P.~J.,   {Lodato} G.,  2022, \mn@doi [\mnras] {10.1093/mnras/stab3442}, \href {https://ui.adsabs.harvard.edu/abs/2022MNRAS.512.2290T} {512, 2290}

\bibitem[\protect\citeauthoryear{{Tsamis}, {Flores-Fajardo}, {Henney}, {Walsh}  \& {Mesa-Delgado}}{{Tsamis} et~al.}{2013}]{2013MNRAS.430.3406T}
{Tsamis} Y.~G.,  {Flores-Fajardo} N.,  {Henney} W.~J.,  {Walsh} J.~R.,   {Mesa-Delgado} A.,  2013, \mn@doi [\mnras] {10.1093/mnras/stt145}, \href {https://ui.adsabs.harvard.edu/abs/2013MNRAS.430.3406T} {430, 3406}

\bibitem[\protect\citeauthoryear{{Vicente}, {Bern{\'e}}, {Tielens}, {Hu{\'e}lamo}, {Pantin}, {Kamp}  \& {Carmona}}{{Vicente} et~al.}{2013}]{Vicente13}
{Vicente} S.,  {Bern{\'e}} O.,  {Tielens} A.~G.~G.~M.,  {Hu{\'e}lamo} N.,  {Pantin} E.,  {Kamp} I.,   {Carmona} A.,  2013, \mn@doi [\apjl] {10.1088/2041-8205/765/2/L38}, \href {https://ui.adsabs.harvard.edu/abs/2013ApJ...765L..38V} {765, L38}

\bibitem[\protect\citeauthoryear{{Villenave} et~al.,}{{Villenave} et~al.}{2020}]{Villenave20}
{Villenave} M.,  et~al., 2020, \mn@doi [\aap] {10.1051/0004-6361/202038087}, \href {https://ui.adsabs.harvard.edu/abs/2020A&A...642A.164V} {642, A164}

\bibitem[\protect\citeauthoryear{{Villenave} et~al.,}{{Villenave} et~al.}{2022}]{Villenave22}
{Villenave} M.,  et~al., 2022, \mn@doi [\apj] {10.3847/1538-4357/ac5fae}, \href {https://ui.adsabs.harvard.edu/abs/2022ApJ...930...11V} {930, 11}

\bibitem[\protect\citeauthoryear{{Wang} \& {Goodman}}{{Wang} \& {Goodman}}{2017}]{Wang17}
{Wang} L.,  {Goodman} J.,  2017, \mn@doi [\apj] {10.3847/1538-4357/aa8726}, \href {https://ui.adsabs.harvard.edu/abs/2017ApJ...847...11W} {847, 11}

\bibitem[\protect\citeauthoryear{{Wilhelm}, {Portegies Zwart}, {Cournoyer-Cloutier}, {Lewis}, {Polak}, {Tran}  \& {Mac Low}}{{Wilhelm} et~al.}{2023}]{Wilhelm23}
{Wilhelm} M. J.~C.,  {Portegies Zwart} S.,  {Cournoyer-Cloutier} C.,  {Lewis} S.~C.,  {Polak} B.,  {Tran} A.,   {Mac Low} M.-M.,  2023, \mn@doi [\mnras] {10.1093/mnras/stad445}, \href {https://ui.adsabs.harvard.edu/abs/2023MNRAS.520.5331W} {520, 5331}

\bibitem[\protect\citeauthoryear{{Winter} \& {Haworth}}{{Winter} \& {Haworth}}{2022}]{Winter23}
{Winter} A.~J.,  {Haworth} T.~J.,  2022, \mn@doi [European Physical Journal Plus] {10.1140/epjp/s13360-022-03314-1}, \href {https://ui.adsabs.harvard.edu/abs/2022EPJP..137.1132W} {137, 1132}

\bibitem[\protect\citeauthoryear{{Winter}, {Clarke}, {Rosotti}, {Hacar}  \& {Alexander}}{{Winter} et~al.}{2019}]{Winter19}
{Winter} A.~J.,  {Clarke} C.~J.,  {Rosotti} G.~P.,  {Hacar} A.,   {Alexander} R.,  2019, \mn@doi [\mnras] {10.1093/mnras/stz2545}, \href {https://ui.adsabs.harvard.edu/abs/2019MNRAS.490.5478W} {490, 5478}

\bibitem[\protect\citeauthoryear{{Winter}, {Haworth}, {Coleman}  \& {Nayakshin}}{{Winter} et~al.}{2022}]{Winter22}
{Winter} A.~J.,  {Haworth} T.~J.,  {Coleman} G. A.~L.,   {Nayakshin} S.,  2022, \mn@doi [\mnras] {10.1093/mnras/stac1564}, \href {https://ui.adsabs.harvard.edu/abs/2022MNRAS.515.4287W} {515, 4287}

\bibitem[\protect\citeauthoryear{{Winter}, {Benisty}, {Manara}  \& {Gupta}}{{Winter} et~al.}{2024}]{2024arXiv240917220W}
{Winter} A.~J.,  {Benisty} M.,  {Manara} C.~F.,   {Gupta} A.,  2024, \mn@doi [arXiv e-prints] {10.48550/arXiv.2409.17220}, \href {https://ui.adsabs.harvard.edu/abs/2024arXiv240917220W} {p. arXiv:2409.17220}

\bibitem[\protect\citeauthoryear{{van Terwisga} \& {Hacar}}{{van Terwisga} \& {Hacar}}{2023}]{VanTerwisga23}
{van Terwisga} S.~E.,  {Hacar} A.,  2023, \mn@doi [\aap] {10.1051/0004-6361/202346135}, \href {https://ui.adsabs.harvard.edu/abs/2023A&A...673L...2V} {673, L2}

\bibitem[\protect\citeauthoryear{{van Terwisga} et~al.,}{{van Terwisga} et~al.}{2020}]{VanTerwsiga2020}
{van Terwisga} S.~E.,  et~al., 2020, \mn@doi [\aap] {10.1051/0004-6361/201937403}, \href {https://ui.adsabs.harvard.edu/abs/2020A&A...640A..27V} {640, A27}

\makeatother
\end{thebibliography}

\label{lastpage}
\end{document}